# Giant gate-controlled odd-parity magnetoresistance in one-dimensional channels with a magnetic proximity effect

Kosuke Takiguchi[1], Le Duc Anh[1,2,3,*], Takahiro Chiba[4], Ryota Fukuzawa[1,5], Takuji Takahashi[5,6] and Masaaki Tanaka[1,6,7,*]

[1] *Department of Electrical Engineering and Information Systems, The University of Tokyo, Bunkyo-ku, Tokyo 113-8656, Japan.*

[2] *Institute of Engineering Innovation, The University of Tokyo, Bunkyo-ku, Tokyo 113-8656, Japan.*

[3] *PRESTO, Japan Science and Technology Agency, Kawaguchi, Saitama, 332-0012, Japan*

[4] *National Institute of Technology, Fukushima College, Iwaki, Fukushima, 970-8034, Japan*

[5] *Institute of Industrial Science, The University of Tokyo, Meguro-ku, Tokyo 153-8505, Japan*

[6] *Institute for Nano Quantum Information Electronics, The University of Tokyo, Meguro-ku, Tokyo 153-8505, Japan*

[7] *Centre for Spintronics Research Network, The University of Tokyo, Bunkyo-ku, Tokyo 113-8656, Japan.*

* Email: anh@cryst.t.u-tokyo.ac.jp
masaaki@ee.t.u-tokyo.ac.jp

**Abstract**

**According to Onsager's principle, electrical resistance *R* of general conductors behaves as an even function of external magnetic field *B*. Only in special circumstances, which involve time reversal symmetry (TRS) broken by ferromagnetism, the odd component of *R* against *B* is observed. This unusual phenomenon, called odd-parity magnetoresistance (OMR), was hitherto subtle (< 2%) and hard to control by external means. Here, we report a giant OMR as large as 27% in edge transport channels of an InAs quantum well, which is magnetized by a proximity effect from an underlying ferromagnetic semiconductor (Ga,Fe)Sb layer. Combining experimental results and theoretical analysis using the linearized**

**Boltzmann's equation, we found that simultaneous breaking of both the TRS by the magnetic proximity effect (MPE) and spatial inversion symmetry (SIS) in the one-dimensional (1D) InAs edge channels is the origin of this giant OMR. We also demonstrated the ability to turn on and off the OMR using electrical gating of either TRS or SIS in the edge channels. These findings provide a deep insight into the 1D semiconducting system with a strong magnetic coupling.**

Investigation of new magnetoresistance (MR) phenomena is an important issue in condensed matter physics, magnetism and spintronics. For example, the discovery of giant MR[1,2] and tunnelling MR[3,4] paved the ways to the creation of non-volatile storage and memory devices. Generally, these MRs are even functions of external magnetic field $B$ according to Onsager's principle[5]. However, it may not be the case when time reversal symmetry (TRS) is broken by magnetism in the system. The odd-parity MR (OMR) in a linear response regime has been observed in systems where TRS is violated.[6,7,8,9] (See also Supplementary Table 1). To explain these OMR phenomena, various possible origins were proposed, including non-trivial Berry curvature, magnetic moments and side jump mechanism[10], and coexistence of spin orbit interaction (SOI) and ferromagnetic coupling in a helical magnet.[11] Even in such rare systems, the OMR magnitude is typically very subtle (the magnitude reported thus far is at most 2%). In addition, these systems reported thus far are metallic, which hinders the control of OMR by external means such as electrical gate voltage.

In this Letter, we report a giant and gate-controlled OMR in the edge transport channels of an InAs thin film interfaced with a ferromagnetic semiconductor (FMS) (Ga,Fe)Sb[12,13,14] layer (see Fig. 1a). The OMR is found to be unprecedently large; the resistance change is 27% of the total resistance when the ***B*** direction is reversed between ± 10 T at $I$ = 1 μA. This is striking, considering that the SOI of InAs is much smaller than other materials such as $SmCo_5$ and pyrochlores in which OMR was observed. We argue that this originates from simultaneous breaking of both TRS and spatial inversion symmetry (SIS), which is entwined with both a Rashba SOI effect at the InAs edges and a strong magnetic proximity effect (MPE) from the underlying (Ga,Fe)Sb.[15] Using field-effect transistor structures, we demonstrate electrical control of the OMR by *individually* tuning the TRS or SIS in the system. The unprecedented strong OMR with gate controllability in mainstream semiconductors such as InAs is ideal not only for elucidating the crucial roles of the TRS and SIS breakings in solid-state physics but also for providing pathways to electronic device applications.

## Results

### Magnetoresistance and its current dependence in InAs/(Ga,Fe)Sb

The structure examined in this study consists of, from top to bottom, InAs (thickness 15 nm)/($Ga_{1-x}$,$Fe_x$)Sb (Fe content $x$ = 20%, 15 nm)/AlSb (300 nm)/AlAs (15 nm)/GaAs (100 nm) on semi-insulating GaAs (001) substrates grown by molecular beam epitaxy (See Fig. 1a). We utilize two samples A and B with the same heterostructure in this study (see Method in detail). In this structure, the InAs layer is a non-magnetic quantum well (QW) that is responsible for over 99% of the electron transport because all the other layers underneath are highly resistive.[15] (Ga,Fe)Sb is a FMS with a high Curie temperature over 300 K.[12,13,14] The preparation and characterization of the samples are explained in Ref. 15. Due to the high crystal quality and staggered band profile at the InAs/(Ga,Fe)Sb interface, in which the conduction band bottom of InAs is at a lower energy than the valence band top of (Ga,Fe)Sb, the electron wavefunction in the InAs QW significantly penetrates into the ferromagnetic (Ga,Fe)Sb layer. This induces a large MPE and spin-dependent scattering in the non-magnetic InAs electron channel.[15]

We pattern the InAs/(Ga,Fe)Sb bilayers into 100 × 600 $\mu m^2$ Hall bars with electrodes labelled '1' to '6', as shown in Fig. 1b. We drive a DC current ***I*** from '1' to '4' and measure the voltage differences $V_{ij} = |V_i - V_j|$ ($i, j$ = 1, 2, 3, 4, 5, 6), from which we obtain the resistances $R_{ij} = V_{ij}/I$. A magnetic field ***B*** is applied perpendicular to the film plane (***B***//$z$). As shown in Fig. 1c and d, the ***B*** dependence of the four-terminal resistance $R_{23}$ measured at $I$ = 1 μA shows i) a very large odd-function MR, ii) a large negative MR, and iii) Shubnikov - de Haas (SdH) oscillations. The last two phenomena ((ii) and (iii)), which are even functions of ***B***, are characteristics of the two-dimensional (2D) electron transport with an MPE in the InAs thin film, as thoroughly discussed in our previous work.[15] Also for (iii), angular dependence of ***B*** also reveals that the SdH oscillations originate from the 2D transport (See Supplementary Fig. 1). In contrast, the large odd-function component, extracted as $R_{23}^{odd}(B)$ (= $[R_{23}(B) - R_{23}(-B)]/2$), is striking. $R_{23}^{odd}$ shows a linear dependence on $B$ with SdH oscillations (see Supplementary Note 1) over the full range of magnetic field ($|B|$ < 10 T) and persists up to 300 K (lower panel of Fig. 1d). $R_{23}^{odd}$ ($B$) is 2.0 kΩ at $B$ = 10 T and 2.5 K, corresponding to 13.5% of the total resistance, and this value changes to 27% upon reversing $B$ to −10 T. This is the largest OMR observed thus far. The OMR magnitude remains almost constant in the whole range of 240 nA < $I$ < 100 μA, drops suddenly to one third of its magnitude at $I_C$ = ~ 200 nA, then remains at this low magnitude when $I$ is decreased further to the lower measurable limit at 50 nA (See Fig. 2a and Supplementary Fig. 3). Even when we reverse the current direction, the OMR remains unchanged (Supplementary Fig. 4). These features indicate that the OMR presented here occurs in a linear transport regime. The reason for the sudden drop at $I_C$ is discussed in Supplementary Note 2. Also, we find that the OMR magnitude depends on the crystallographic orientation, which may be due to the non-uniformity of Fe atoms in (Ga,Fe)Sb (See Supplementary Note 3).

**One-dimensional (1D) transport in InAs and the origin of OMR**

An important observation, obtained by comparing $R_{23}$ and $R_{65}$ in Fig. 2b, is that the sign of the OMR flips when we switch the voltage terminals contacting the side edge while maintaining the same measurement setup. Given that $\boldsymbol{B}$ and $\boldsymbol{I}$ are fixed in the same directions, this observation suggests that the OMR originates from the electrical transport along the side edges of the InAs thin film, where the SIS is broken by the opposite polarities, as discussed in the next paragraph. This argument is further supported by the disappearance of the OMR in our two-terminal resistance measured between the electrodes 1 and 4 ($R_{14}$), where the positive and negative OMR components from the two side edges of the InAs thin film exactly cancel (see Fig. 2c, and Supplementary Note 4). We note that, however, a large OMR was observed when we measured the resistance only along one edge by the two-terminal method (see Supplementary Note 5).

Two types of edge transport are known to occur in InAs/GaSb bilayers. One involves a non-trivial quantum spin Hall edge state,[16,17,18] which is formed at the edge of the InAs/GaSb interface when a topological gap is opened due to the inverted band structure (the valence band top of GaSb is at a higher energy than the conduction band bottom of InAs) and SOI. However, because this topological gap is very small (~4 meV), the non-trivial edge state cannot survive at high temperature, which contradicts our observation of the OMR up to room temperature. The other involves a trivial edge state formed at the edge of the InAs layer due to the pinning of the Fermi level at the top and side vacuum surfaces, which is located as high as 0.1 - 0.3 eV above the conduction band bottom.[19,20,21,22,23,24] As a result, the conduction band potential of InAs is strongly bent downward at the surfaces, which we confirmed using Kelvin force microscopy measurements (See Methods and Supplementary Fig. 10). The effect is two-fold: First, the electron carriers accumulate more at the edges than in the centre of the InAs film; thus, two 1D edge channels and one 2D transport channel coexist. This fact is confirmed by the transport measurements on devices with different sizes, which is discussed in Supplementary Note 6. Second, the SIS is broken at the side edges due to the resulting built-in electric field. Since we define the directions of $\boldsymbol{I}$ and $\boldsymbol{B}$ in our measurements as the $x$ and $z$ directions, respectively, as shown in Fig. 1a, the built-in electric field $\boldsymbol{E}_{\mathrm{sur}}$ points outward along the $y$ direction. The directions of $\boldsymbol{E}_{\mathrm{sur}}$ in the two edge channels are opposite, which explains the opposite signs of the OMRs in $R_{23}$ and $R_{65}$. As shown in Fig. 2d, the OMR almost disappears when we apply $\boldsymbol{B}$ parallel to the current $\boldsymbol{I}$ direction (the $x$ axis) or the $\boldsymbol{E}_{\mathrm{sur}}$ direction (the $y$ axis) (see Supplementary Fig. 12). This indicates that OMR can only be induced when $\boldsymbol{B}$, $\boldsymbol{I}$, and $\boldsymbol{E}_{\mathrm{sur}}$ are mutually orthogonal. This is also because the MPE from (Ga,Fe)Sb, which breaks the TRS in InAs, is only effectively induced by the $z$-component of the magnetization of (Ga,Fe)Sb[15].

**Control of SIS and TRS breaking via gate voltage**

To examine our scenario, we apply electrical gate voltage to *individually* tune the TRS and SIS breakings in the edge transport of InAs and evaluate their impacts on the OMR. We fabricated two field-effect transistor devices D1 and D2; one (D1) has a single gate electrode G that controls the whole InAs Hall bar (Fig. 3a), and the other (D2) has two separated gate electrodes $G_1$ and $G_2$ that control the conduction of each edge independently (Fig. 3b). In device D1, a negative (positive) voltage applied to G push the electron wavefunctions in InAs towards the (Ga,Fe)Sb (top surface) side, which effectively enhances (suppresses) the MPE[15]. As shown in Fig. 3c, in device D1, when applying a negative gate voltage $V_g$ from 0 V to −5 V on G, with which the MPE is enhanced, the OMR intensity strongly increases by more than three folds (2.5% to 8%, respectively). Meanwhile, when applying a positive $V_g$ from 0 V to 5 V on G, which effectively suppresses the MPE, the OMR intensity decreases and almost vanishes at $V_g$ = −5 V. These results clearly demonstrate the important role of TRS breaking in inducing the OMR. This fact is also confirmed by the small OMR magnitude (= 1.8% at 14 T) in an InAs/GaSb reference sample, where there is no FM coupling, as shown in Supplementary Fig. 13. On the other hand, in device D2, by applying a voltage in one of these two gates (for example, G1), we modulate the band profile in one edge of the InAs layer (the edge along terminals 2 and 3). This enhances the OMR in one edge than another, and results in an appearance of OMR *even* in the magnetoresistance measured between the terminals 1 and 4. Figure 3d shows the magnetoresistance characteristics measured between terminals 1 and 4 when we applied $V_{g1}$ = 7 V and −7 V on G1. One can see that a negative (positive) OMR is induced at $V_{g1}$ = 7 V (−7 V) as expected. This can be understood because a positive (negative) $V_{g1}$ enhances (suppresses) the $\boldsymbol{E}_{sur}$ of the right edge in relative to that of the left edge. Therefore, the important role of SIS breaking at the edge channels is clearly demonstrated by these results.

**Theoretical analysis**

Finally, we discuss the theoretical model to explain the OMR in InAs/(Ga,Fe)Sb.

If we temporarily neglect the MPE from the (Ga,Fe)Sb layer, the Hamiltonian of the 1D edge channel of InAs can be described as

$$\hat{H}_{1\mathrm{D}}(k_x) = \frac{\hbar^2 k_x^2}{2m^*}\sigma_0 + (\Lambda_{\mathrm{side}} k_x + \Delta_z)\sigma_z + \Lambda_{\mathrm{top}} k_x \sigma_y \quad (1)$$

where $k_x$ is the wavenumber along the $x$ direction, $m^*$ is the effective mass of electrons, $\Lambda_{\mathrm{top(side)}}$ (= $\hbar\lambda_{\mathrm{top(side)}}$) is the effective Rashba SOI due to the built-in potential at the top (side edge) surface, $\hbar$ is the Dirac's constant, $\Delta_z$ (= $g\mu_B B_z$) is the Zeeman splitting due to an applied magnetic field along the $z$-axis ($B_z$), $\sigma_i$ ($i = x, y, z$) are the elements of the Pauli matrix that acts on the electron spin degree of freedom, and $\sigma_0$ is the identity matrix. The energy dispersion from eq. (1) can be described as

$$E_s = \frac{\hbar^2 k^2}{2m^*} + s\sqrt{(\Lambda_{\mathrm{side}} k + \Delta_z)^2 + \left(\Lambda_{\mathrm{top}} k\right)^2} \quad (2)$$

where $s$ = +/− denotes the upper and lower bands $E_+$ and $E_-$, as depicted in Fig. 4a, respectively. Here we define the energy band bottom of $E_-$ as $E = 0$. It is important to note that due to the Rashba SOI ($\Lambda_{\mathrm{top}}$ and $\Lambda_{\mathrm{side}}$), the spin components $\sigma_y$ and $\sigma_z$ are locked to the momentum $k_x$ in opposite directions between the bands $E_+$ and $E_-$. Thus the + and – subscripts also indicate the difference of "chirality" of these bands, which are shown as green and pink lines, respectively, in the right-side graph of Fig. 4a. We solve Boltzmann's equations and obtain the electrical conductivity $\sigma_{xx}$ by summing the conductivities of all the bands that cross the Fermi level ($E_F$) (see Supplementary Note 7),

$$\sigma_{xx} \simeq \frac{e^2}{h}\sum_s \tau_s \int dE_s \sqrt{1 + \frac{2E_s}{m^*\lambda_{\mathrm{side}}^2}}\left(1 - s\frac{|\lambda_{\mathrm{side}}|}{\lambda_{\mathrm{side}}}\frac{\Delta_z}{m^*\lambda}\right)\delta(E_s - E_F) \quad (3)$$

where $e$ is the elementary charge, $\tau_s$ is the relaxation time, $h$ is the Planck's constant, and $E_F$ is the Fermi energy. Reflecting breaking of the SIS at the side surface edges, we assume $\Lambda_{\mathrm{top}} \ll \Lambda_{\mathrm{side}}$, which indicates that the electric field at the side edges is much larger than that at the top surface.[25] From eq. (3) and Fig. 4a, the odd-order $B_z$-dependent conductivity can be non-zero in the case that $E_F$ crosses only the lower band shown in region (II) of Fig. 4a. However, this case is unlikely because the gap $\Delta_g(B)$ is only 24 meV and 44 meV at $B$ = 0 and 14 T, respectively, obtained by using the parameters of an InAs nanowire of $m^*/m_0 = 0.08$,[26] $g = 18$,[27] $m^*\lambda^2_{side} = 0.45$ meV[26], and $m^*\lambda^2_{top} = 0.027$ meV.[28] Due to the Fermi level pinning at the edge surface, $E_F$ in the edge channel lies in region (I) of Fig. 4a where the odd-order $B_z$-dependent conductivities from the upper and lower bands cancel each other out, and thus, no OMR should be expected.

However, the OMR can be induced if the relaxation times in the $E_+$ and $E_-$ bands are different ($\tau_+ \neq \tau_-$), which results from the MPE and the Rashba SOI in the 2D and 1D channels of InAs as explained in the following. Our analysis of the transport data (see Supplementary Note 2 and 6) indicates that the MPE mainly affects the 2D channel, inducing a splitting energy gap $\Delta_{2D}$ between 2D bands of opposite $\sigma_z$ components (indicated by purple arrows in in Fig. 4b). Therefore, the MPE affects the 1D channel only indirectly via electron scattering between the 1D and 2D channels. Considering that $\sigma_y$ is locked to $k_x$ because of $\Lambda_{top}$ in both the 1D and 2D channels, the lower and upper bands in each channel (1D and 2D) have different chiralities, as indicated by the green and pink colours in Fig. 4b. The relaxation time of $E_+$ and $E_-$ ($\tau_+$ and $\tau_-$, respectively) in the 1D channel thus are mainly determined by scattering events between bands with the same chirality (blue and red arrows in Fig. 4b). The difference in the density of states (DOS) at the Fermi level of the two 2D bands (pink and green bands in Fig. 4b) then leads to the asymmetric scattering between $E_+$ and $E_-$ in the 1D edge channel, and different values of $\tau_+$ and $\tau_-$ (see Supplementary Note 8 for detailed discussions). Consequently, the linear-response conductivity $\sigma_{xx}$ is rewritten as

$$\sigma_{xx} \simeq \frac{e^2}{h}\tau_-|\lambda_{\text{side}}|\sqrt{1+\frac{2E_F}{m^*\lambda_{\text{side}}^2}}\left[1+\alpha+(1-\alpha)\frac{|\lambda_{\text{side}}|}{\lambda_{\text{side}}}\frac{g\mu_B}{2E_F+m^*\lambda_{\text{side}}^2}B_z\right] \quad (4)$$

Here, we set the phenomenological parameter $\alpha$ as $\tau_+ = \alpha\tau_-$ to express the different relaxation time of electron carriers in the $E_+$ and $E_-$ states. Under the influence of the strong MPE and chirality-dependent scattering at the interface ($\alpha \ll 1$), the linear $B_z$-dependent MR appears in the conductivity $\sigma_{xx}$ due to the contribution of the last term in the brackets of eq. (4). Using $\alpha = 0.1$, $E_F = 100$ meV, $m^*\lambda^2_{\text{side}} = 0.45$ meV,[26] and $\Delta_z/B_z = 0.52$ meV/T for $g = 18$,[28,] the OMR is clearly reproduced by eq. (4), as shown in Fig. 4c. The different signs of $R_{23}^{\text{odd}}$ and $R_{65}^{\text{odd}}$ shown in Fig. 2b are explained by the different signs of the Rashba parameter $\lambda_{\text{side}}$ (blue and red lines) between the two side edges. The dependences of the OMR ratio $\Delta R/R_0$ on $\alpha$ and $E_F$ are shown in Fig. 4d and e, respectively. A large difference in the relaxation time of the spin channels, which means a small $\alpha$, produces a large OMR ratio. This indicates the important role of MPE at the InAs/(Ga,Fe)Sb interface in inducing the large OMR. This conclusion is also supported by the fact that the OMR magnitude $\Delta R/R_0$, where $\Delta R = R_{23}^{\text{odd}}(10\text{ T})$ and $R_0 = R_{23}(0\text{ T})$, is enhanced with decreasing temperature $T$ as $\ln(1/T)$ (see the inset of Fig. 1d). This behaviour is characteristic of the Kondo-effect-related transport coming from the spin-dependent scattering at the InAs/(Ga,Fe)Sb interface. Another important result is that a smaller $E_F$ leads to a larger OMR. If we set $E_F$ at approximately 24 meV, which is the same as $\Delta_g(0\text{ T})$, then eq. (4) can reproduce the experimental value ($\Delta R/R_0 = 13.5\%$), as shown in Fig. 4e.

In conclusion, we found the giant odd-parity magnetoresistance in the 1D edge channels of the InAs/(Ga,Fe)Sb heterostructure, and demonstrated the ability to electrically turn on and off the effect using field-effect transistor structures. Our results highlight the abundance of new physics in solid state systems when TRS and SIS are simultaneously broken even in well-known materials such as InAs. The linear OMR presented in this work can be applied to magnetic field sensors which provides a large dynamic range (0 − 10 T) owing to its linearity. This new type of sensors can work at room temperature, requires only simple DC measurements for detection, and its sensitivity can be further enhanced by material engineering, such as optimizing the carrier concentration and SOI strength.

**Methods:**

**Sample preparation and characterization**

We grew heterostructures consisting of InAs (thickness 15 nm)/(Ga,Fe)Sb (15 nm, Fe 20%, $T_C$ > 300 K)/AlSb (300 nm)/AlAs (15 nm)/GaAs (100 nm) on semi-insulating GaAs (001) substrates by molecular beam epitaxy (MBE). The growth temperature ($T_S$) was 550°C for the GaAs and AlAs layers, 470°C for the AlSb layer, 250°C for the (Ga,Fe)Sb layer, and 235°C for the InAs layer. We also grew a nonmagnetic InAs/GaSb heterostructure as a reference, whose structure is the same as the sample mentioned above, except for the lack of Fe doping. The top two layers (InAs and GaSb) of this sample were grown at 470°C, while the other layers were grown under the same conditions as the Fe-doped samples. The *in situ* reflection high energy electron diffraction (RHEED) patterns of InAs and (Ga,Fe)Sb are bright and streaky, indicating good crystal quality and a smooth surface (see Supplementary Fig. S2b in Ref. 15). In this paper, we used two different samples A and B of InAs/(Ga,Fe)Sb heterostructures with sheet carrier concentrations 2.0 × $10^{12}$ $cm^{-2}$ and 1.8 × $10^{12}$ $cm^{-2}$, and electron mobilities 9.4 × $10^2$ $cm^2$/Vs and = 1.9 × $10^3$ $cm^2$/Vs, respectively. Also, the quantum mobility of sample A is estimated to be 2070 $cm^2$/Vs from the SdH oscillations (See Supplementary Fig. 14).

The mobility difference between sample A and B suggests that the static electric fields, which determine the confinement potential in the 1D and 2D channels of InAs, in the two samples are different. The confinement potential sensitively affects the strength of both the proximity magnetoresistance[15] and the Rashba SOI in the InAs channel, and consequently yields different OMR in these sample A and B. We note that the different confinement potentials may originate from different surface pinning effects at the top and side surfaces of the devices, which depend on the detailed conditions during device fabrication.

**Fabrication process of the Hall bar devices and transport measurement**

The samples were patterned into 100 × 600 $\mu m^2$ Hall bars using standard photolithography and Ar ion milling down to the AlSb buffer layer. The etched surface was passivated by depositing a thin $SiO_2$ layer. Then electrodes were formed by electron-beam evaporation and lift-off of Au (50 nm)/Cr (5 nm) films. Figure 1b shows an optical microscopy image of the Hall bar device examined in Figs. 1, 2, and 3c. For the field-effect transistor (FET) devices in Fig. 3c and d, we deposited a 50 nm-thick $Al_2O_3$ layer as a gate insulator by atomic layer deposition. Figure 3a and b show optical microscopy images of the Hall bar FET device examined in Fig. 3c (D1) and d (D2), respectively. Magnetotransport measurements were conducted using a Quantum Design physical property measurement system (PPMS) by a standard 4-terminal method, except for $R_{14}$ which was measured by a 2-terminal method. We use a DC current for $I > 1\ \mu$A, and an AC current with lock-in amplifier (lock-in frequency is 5261 Hz) for lower $I$.

**Work function measurements by Kelvin probe force microscopy (KFM)**

We investigated a distribution of the surface potential on the InAs/(Ga,Fe)Sb by KFM in vacuum condition (~ $10^{-5}$ Pa) at room temperature. In KFM, an AC bias at frequency $f$ (= 1 kHz in our case) and a DC bias are applied between the tip and the sample under noncontact operation in atomic force microscopy (AFM) (see Supplementary Fig. 10 in Supplementary Information). When the tip approaches the sample surface in the $z$ direction, the electric bias induces an electrostatic force $F$ expressed as

$$F = \frac{1}{2}\frac{dC}{dz}\left(V_{\mathrm{dc}} - \frac{\Delta\phi}{e} + V_{\mathrm{ac}}\sin 2\pi f t\right)^2$$

$$= \frac{1}{2}\frac{dC}{dz}\left(V_{\mathrm{dc}} - \frac{\Delta\phi}{e}\right)^2 + \frac{1}{4}\frac{dC}{dz}V_{\mathrm{ac}}^2 + \frac{dC}{dz}\left(V_{\mathrm{dc}} - \frac{\Delta\phi}{e}\right)V_{\mathrm{ac}}\sin 2\pi f t - \frac{1}{4}\frac{dC}{dz}V_{\mathrm{ac}}^2\cos 4\pi f t, \quad (6)$$

where $C$ is a capacitance between tip and the sample, $V_{\mathrm{dc}}$ is the DC bias voltage, $V_{\mathrm{ac}}$ is the AC voltage magnitude, and $\Delta\phi$ is the work function difference between the tip and the sample. Similar to AFM measurements, the force $F$ is deduced from the shift of the cantilever oscillation frequency. $V_{\mathrm{dc}}$ is adjusted using a feed-back control so that the $f$-frequency component in $F$, which is measured using a lock-in amplifier, is nullified. Then $V_{dc}$ gives the value of $\Delta\phi/e$ according to eq. (6). Therefore, we can obtain $\Delta\phi$ and consequently the work function distribution on the sample.

We note that, the potential profile at the topmost InAs surface detected by KFM might be different from the one at several-atomic-layer depth below the surface. This is because of a screening effect from a large amount of charged surface states (top and side surfaces), which are common at InAs surfaces. Thus, the potential profile change along the $y$ direction measured by the KFM tip is much *milder* than the real confinement potential at the edge of the InAs channel.[29] As a result, the potential profile at the top surface shown in Supplementary Fig. 10b might largely exaggerate the width of the triangular potential at the bulk InAs side surface, which should be much less than 2 μm. Therefore, we consider that the static electric field in the *y* and *z* directions should have same order of magnitude.

**Acknowledgments:** A part of this work was conducted at Advanced Characterization Nanotechnology Platform of the University of Tokyo, supported by "Nanotechnology Platform" of the Ministry of Education, Culture, Sports, Science and Technology (MEXT), Japan. K.T. acknowledges the support from the Japan Society for the Promotion of Science (JSPS) Fellowships for Young Scientists. K. T. and R. F. acknowledge the Material Education program for the future leaders in Research, Industry, and Technology (MERIT).

**Funding:** This work was partly supported by Grants-in-Aid for Scientific Research (Nos. 16H02095, 17H04922, 18H05345, 20H05650), the CREST Program (JPMJCR1777) and PRESTO Program (JPMJPR19LB) of the Japan Science and Technology Agency, and the Spintronics Research Network of Japan (Spin-RNJ).

**Author contributions:** K. T. and L. D. A. designed the experiments and grew the samples. K. T. fabricated devices, performed sample characterizations and transport measurements. R. F. and T. T. conducted KFM measurements. K. T., L. D. A, T. C. and M. T. discussed on the mechanism. T. C. performed theoretical calculations. K. T., L. D. A. and M. T. wrote the manuscript. L. D. A. and M. T. supervised the study.

**Competing interests:** Authors declare no competing interests.

**Data and materials availability:** All data are available in the main text or the Supplementary Information.

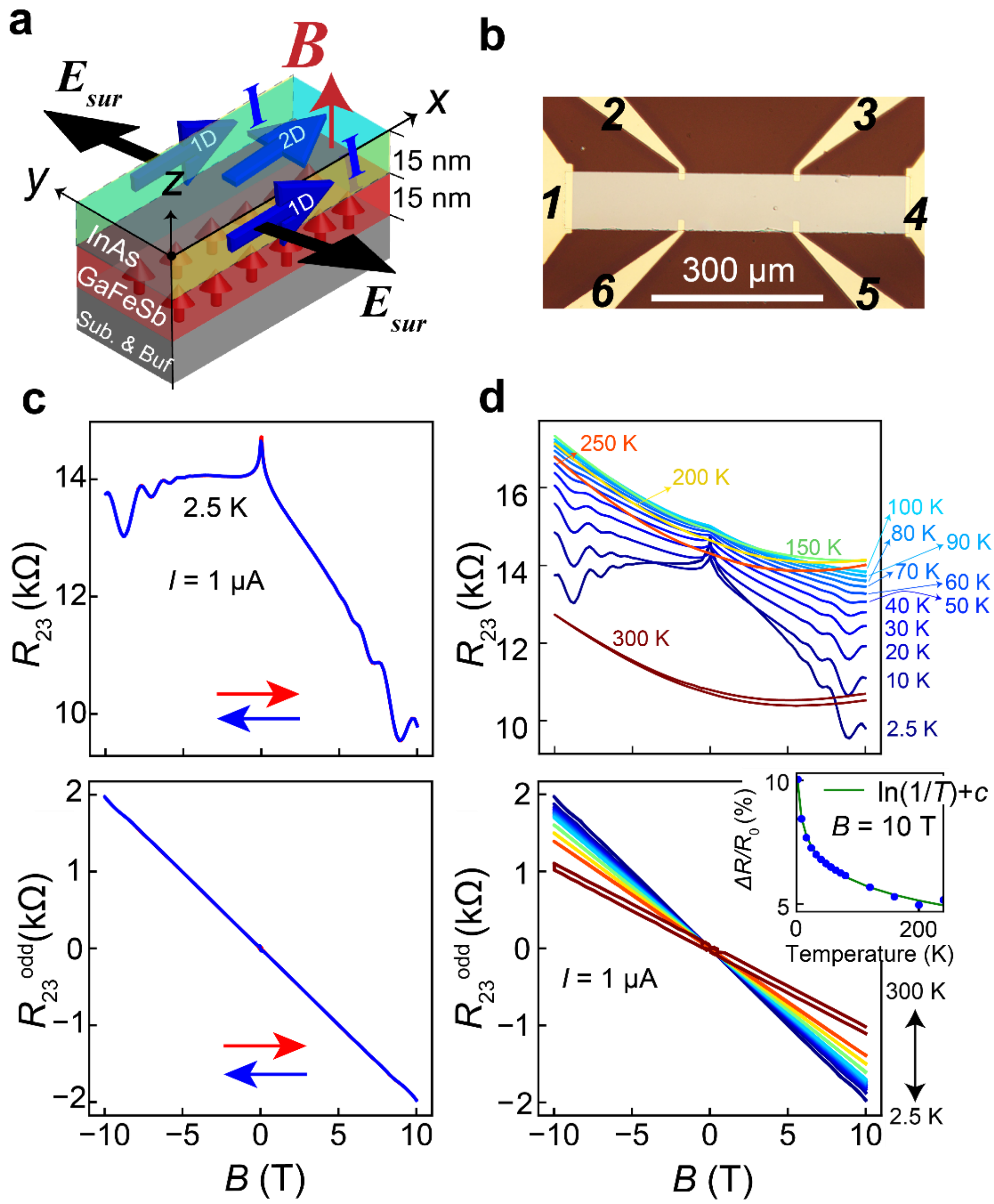

**Fig. 1| Magnetoresistances (MRs) of InAs/(Ga,Fe)Sb bilayer heterotructures. a,** Schematic illustration of the InAs/(Ga,Fe)Sb heterostructure with 1D transport channels at the side edges. We applied an electric current ***I*** parallel to the *x* direction and an external magnetic field ***B*** parallel to the *z*-axis. Because (Ga,Fe)Sb is insulating, electron carriers flow only in the InAs QW layer, both in the 2D channel and the 1D channels at the edges. The triangular potentials at the side surfaces create static electric fields $\boldsymbol{E}_{\mathbf{sur}}$ parallel to the *y*-axis at the side edges of the InAs QW. **b,** Optical microscopy top view image of the device. The terminals are labelled '1' – '6', as shown in the image. **c**, (Upper panel) MR of the InAs/(Ga,Fe)Sb heterostructure of sample A, measured with a DC current of 1 μA

and an external magnetic field ***B*** applied parallel to $z$ at 2.5 K. The blue and red arrows indicate the sweep direction of ***B***. (Bottom panel) Extracted odd components of the upper panel data ($R_{23}^{\mathrm{odd}} = [R_{23}(B) - R_{23}(-B)]/2$). **d**, Temperature dependences of $R_{23}$ and $R_{23}^{\mathrm{odd}}$ of sample A at 2.5 − 300 K with $I$ = 1 μA. Although the even-function MR and the Shubnikov-de Haas oscillation disappear at high temperature, the OMR component remains up to 300 K. The inset of the lower panel shows the temperature dependence of $\Delta R/R_0$, where $\Delta R = R_{23}^{\mathrm{odd}}(10\ \mathrm{T})$ and $R_0 = R_{23}(0\ \mathrm{T})$ (blue circles). The green curve is the fitting result obtained using the logarithmic function $\ln(1/T) + c$ ($T$, temperature; $c$, temperature-independent parameter).

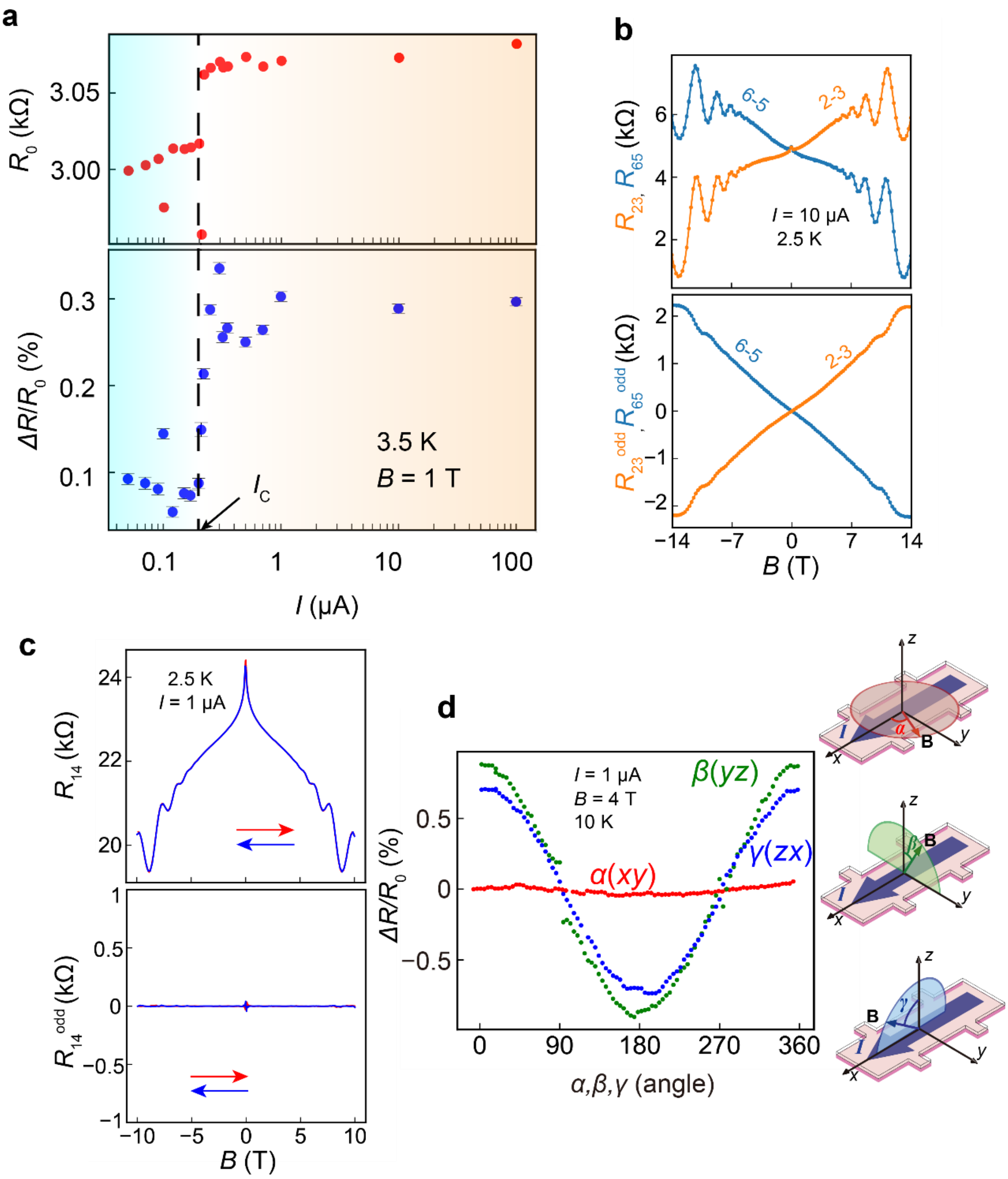

**Fig. 2| Properties of the OMR in 1D InAs edge channels. a,** Current $I$ dependence of the zero field resistance $R_0$ (upper panel) and the OMR magnitude $\Delta R/R_0$ (= $[(R_{23}(1\ \mathrm{T}) - R_{23}(-1\ \mathrm{T}))/2]/R_0$) (lower panel) measured in $R_{23}$ of sample B at 3.5 K. The $R_0$ and $\Delta R/R_0$ jump up at $I_C$ (= 200 nA) simultaneously. **b**, Comparison of the $B$-dependences of $R_{23}$ and $R_{65}$ (upper panel) of sample A and their odd-function components (lower panel) measured with a fixed current of 10 μA at 2.5 K. $R_{23}$ and $R_{65}$, which are measured along the different 1D channels at the opposite edges, show opposite $B$ dependences. **c**, MR

curve of $R_{14}$ of sample A (upper panel) and its odd component (lower panel) measured with a fixed current of 1 μA at 2.5 K. The OMRs in the opposite 1D channels cancel each other out, leading to an almost zero odd component in $R_{14}$. **d**, Angle dependences of the OMR magnitude $\Delta R/R_0$ of sample B, where $\Delta R = (R_{23}(4\ \mathrm{T}) - R_{23}(-4\mathrm{T}))/2$ and $R_0 = R_{23}(0\ \mathrm{T})$. The red, green and blue dots indicate the OMR magnitude in the *xy*, *yz* and *zx* rotations, respectively.

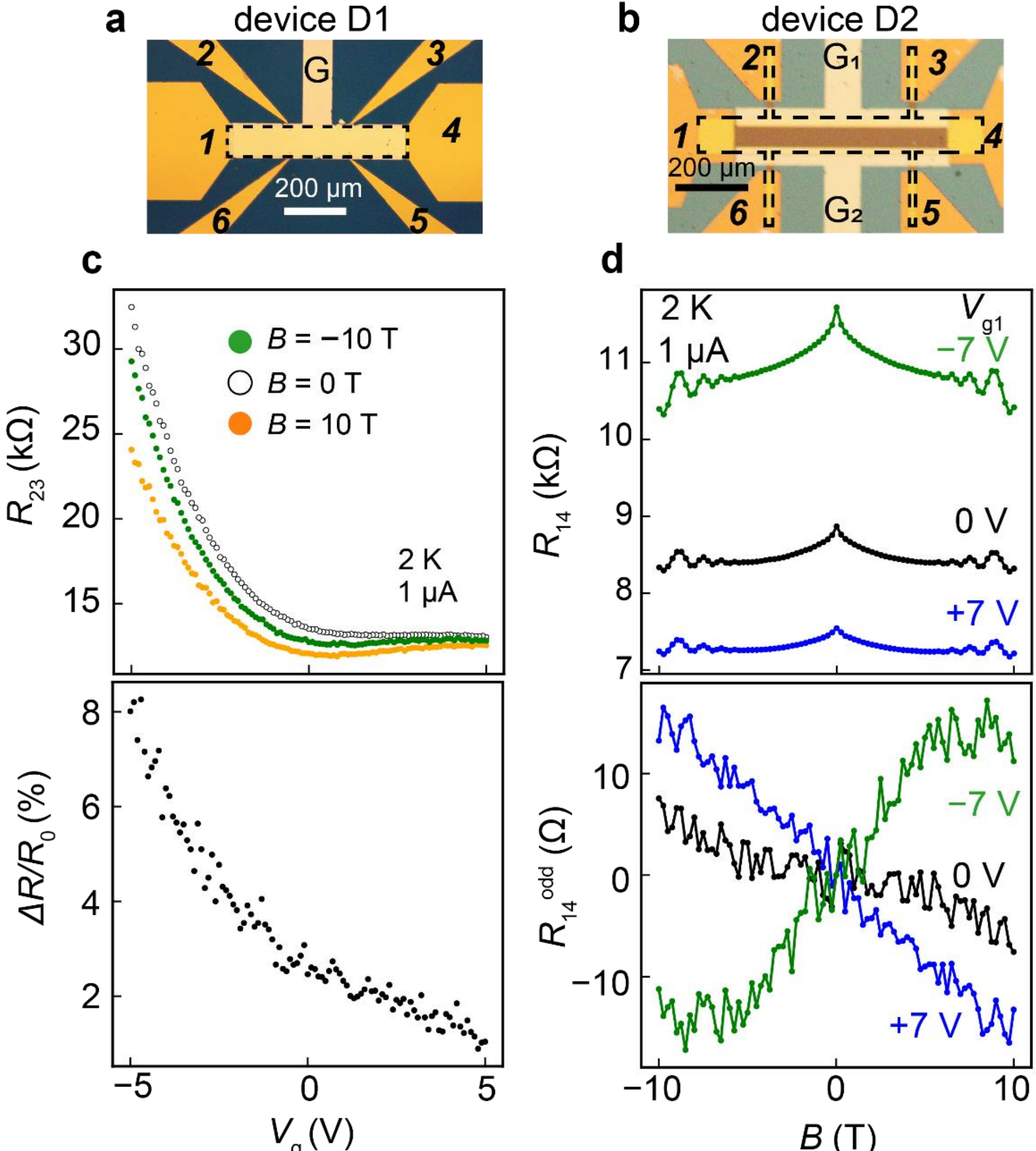

**Fig. 3| Electrical gating of the OMR. a**, **b**, Optical microscopy image of the Hall-bar field-effect transistor (FET) device D1 with a top-gate G and D2 with separated gates $G_1$ and $G_2$, respectively. These two devices are made from sample B. The dashed line indicates the outline of the Hall bars. The light and dark yellow parts are the Au pads of gate and Hall-bar electrodes, respectively. **c**, Gate voltage $V_g$ dependence of $R_{23}$ at various $B$ of −10 T (green), 0 T (white), 10 T (orange) (upper panel), and that of the odd component $\Delta R/R_0$ (lower panel), where $\Delta R = (R_{23}(-10\text{ T}) - R_{23}(+10\text{ T}))/2$, and $R_0 = R_{23}(0\text{ T})$, measured at 2 K on device D1. These measurements are conducted with a fixed current of 1 μA at 2 K. **d**, Magnetoresistance (MR) results of (top panel) and the odd components (bottom panel), measured at 2 K on device D1. $R_{14}$ is the resistance measured between terminals 1 and 4 (two-terminal measurement). The MR results at $V_{g1}$ = −7, 0, +7 V are shown in green, black and blue lines, respectively.

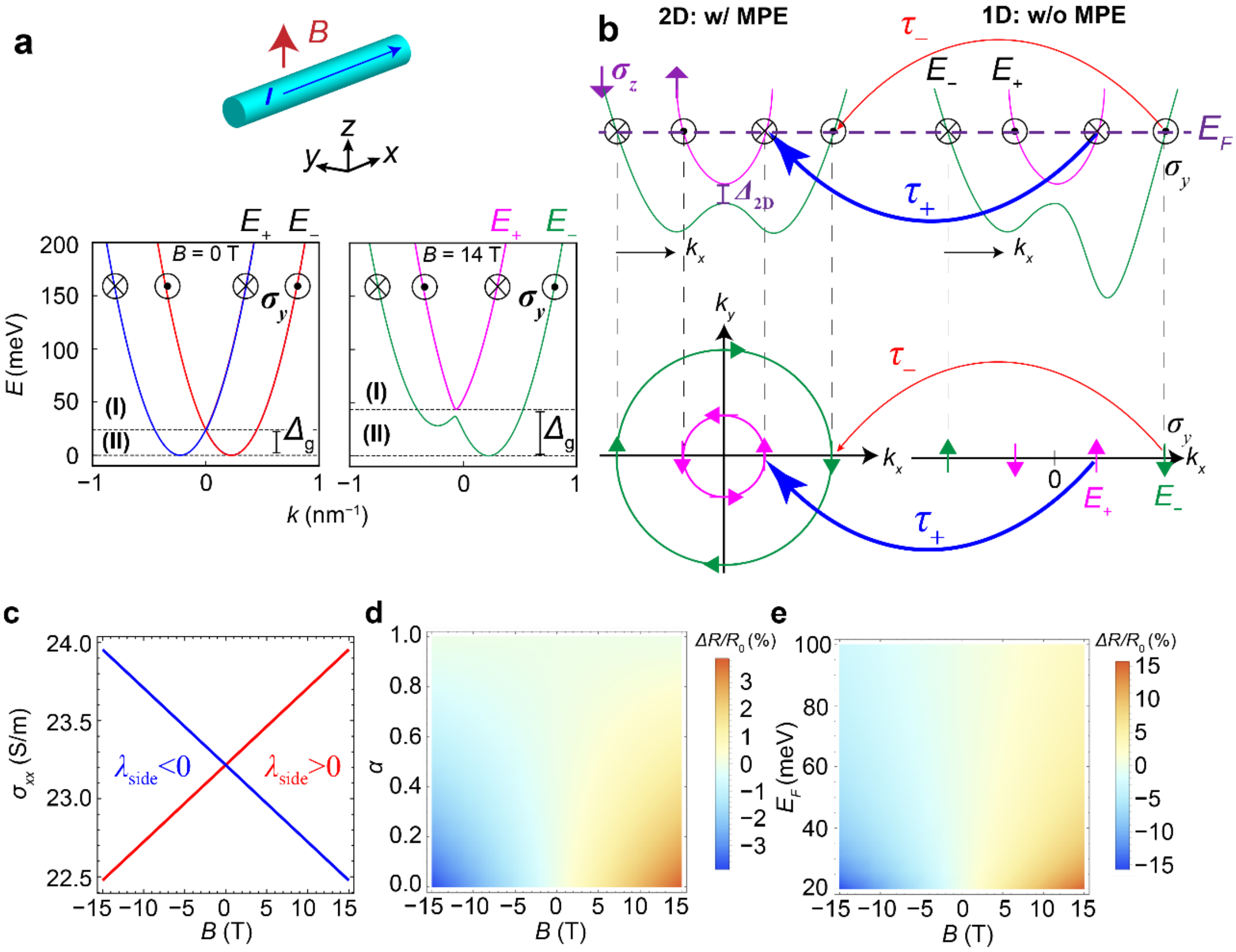

**Fig. 4| Theoretical calculation using Boltzmann's equation in the 1D Rashba system with an MPE. a,** Energy band dispersions ($E_+$, $E_-$) of the 1D Rashba system under a magnetic field $B$ (0 T and 14 T, shown in the left and right panels, respectively) applied parallel to the $z$ direction, calculated by eq. (2). Here, we set $m^*/m_0 = 0.08$,[26] $\Delta_z/B_z = 0.52$ meV/T for $g = 18$,[27] $m^*\lambda^2_{\text{side}} = 0.45$ meV,[28] and $m^*\lambda^2_{\text{top}} = 0.027$ meV.[28] In the case of $B$ = 14 T, $E_+$ and $E_-$ can be labelled by chirality (pink and green lines, respectively) which is determined by the SIS breaking in the $z$ direction. Two different regions (I) and (II) can be observed, defined by whether the Fermi energy $E_F$ crosses only one or two dispersion branches $E_+$ and $E_-$. $\Delta_g$ is the energy gap between the minima of $E_+$ and $E_-$. **b**, Schematic energy dispersion (upper) and its Fermi surface (lower) of the 1D (right) and 2D (left) channels. Purple dashed line indicates the Fermi level $E_F$. Due to the Rashba SOI in the $z$ direction in both channels, the $y$ spin component ($\sigma_y$) of electrons is locked to the momentum $k_x$ in opposite directions between the green and pink bands. Electron scattering between the 1D and 2D channels occurs mainly between bands with the same chirality, as indicated by the red and blue arrows. In the 2D channel, MPE opens the gap

($\Delta_{2D}$) between two bands with opposite $z$ spin components ($\sigma_z$). Different density of states between the two (pink and green) 2D bands leads to different relaxation times $\tau_+$ and $\tau_-$ in the 1D channel. **c**, Calculated results of the OMR using eq. (4) with $\alpha = 0.1$, $E_F = 100$ meV, $m^*\lambda^2_{side} = 0.45$ meV,[28] and $\Delta_z/B_z = 0.52$ meV/T for $g = 18$.[27] The sign of the Rashba parameter $\lambda_{side}$ determines the polarity of the OMR component in the 1D system. **d,** and **e,** OMR as functions of $\alpha$ (with $E_F = 100$ meV) and $E_F$ (with $\alpha = 0.1$), respectively. $\alpha$ represents the strength of the MPE at the interface; $\alpha$ is small in the case of a strong MPE. A strong MPE and a small $\alpha$ lead to a large OMR.

# Supplementary Information

# Giant gate-controlled odd-parity magnetoresistance in one-dimensional channels with a magnetic proximity effect

Kosuke Takiguchi[1], Le Duc Anh[1,2,3,*], Takahiro Chiba[4], Ryota Fukuzawa[1,5], Takuji Takahashi[5,6] and Masaaki Tanaka[1,6,7,*]

[1] *Department of Electrical Engineering and Information Systems, The University of Tokyo, Bunkyo-ku, Tokyo 113-8656, Japan.*
[2] *Institute of Engineering Innovation, The University of Tokyo, Bunkyo-ku, Tokyo 113-8656, Japan.*
[3] *PRESTO, Japan Science and Technology Agency, Kawaguchi, Saitama, 332-0012, Japan*
[4] *National Institute of Technology, Fukushima College,Iwaki, Fukushima, 970-8034, Japan*
[5] *Institute of Industrial Science, The University of Tokyo, Meguro-ku, Tokyo 153-8505, Japan*
[6] *Institute for Nano Quantum Information Electronics, The University of Tokyo, Meguro-ku, Tokyo 153-8505, Japan*
[7] *Centre for Spintronics Research Network, The University of Tokyo, Bunkyo-ku, Tokyo 113-8656, Japan.*

## Supplementary Note 1: Oscillating behaviour of the OMR

As shown in Fig. 2b, the odd-parity magnetoresistance (OMR) curves show oscillating behaviour. This phenomenon can be explained by Landau quantization. In our theoretical model, Boltzmann's equation describes the OMR in a 1D case as follows.

$$\sigma_{xx} \simeq \frac{e^2}{h}\tau_-|\lambda_{\mathrm{side}}|\sqrt{1+\frac{2E_F}{m^*\lambda_{\mathrm{side}}^2}}\left[1+\alpha+(1-\alpha)\frac{|\lambda_{\mathrm{side}}|}{\lambda_{\mathrm{side}}}\frac{g\mu_B}{2E_F+m^*\lambda_{\mathrm{side}}^2}B_z\right] \quad (3)$$

In this equation, the electrical conductivity $\sigma_{xx}$ is proportional to the relaxation time of $\tau_-$, where the subscript "–" indicates that $\tau_-$ is the relaxation time of the lower band $E_-$ (see Fig. 4a).

Generally, the external magnetic field quantizes the density of states (DOS) (Landau quantization), leading to the quantum (SdH) oscillation in the $\sigma_{xx} - B$ characteristics. Since the relaxation time $\tau$ is proportional to DOS, $\tau$ can be described as

$$\frac{1}{\tau(E,B)} = \frac{1}{\tau_0}\left(1+\frac{\Delta D}{D_0}\right) \quad (\mathrm{S1})$$

where $\tau_0$ represents the relaxation time that is independent of the electric field $E$ and magnetic field $B$, and $\Delta D/D_0$ represents the $B$-dependent oscillation component of DOS. Since $\tau_-$ is obtained from eq. (S1), via this relaxation time $\tau(E,B)$, the Landau quantization can manifest itself as oscillation in the 1D transport and the OMR.

According to the Lifshitz-Kosevich theory,[S1,S2] the quantum oscillation becomes clear when the coherence length is long and the mobility is high. This is indeed confirmed in our new sample with higher mobility (= $1.9\times10^3$ cm$^2$/Vs) than the previous sample (= $9.4\times10^2$ cm$^2$/Vs), as shown in Supplementary Fig. 2. The odd component exhibits much clearer oscillation than the previous sample, which supports our conclusion that the oscillation in OMR originates from the Landau quantization.

## Supplementary Note 2: Possible origin of the current dependence of the OMR around $I_C$ = 200 nA

In the InAs channel, there are parallel conductions in the edge (one dimensional (1D)) and center (two-dimensional (2D)) channels (see Supplementary Fig. 5a). In both 1D and 2D channels, there are magnetic proximity effects (MPE) induced by the perpendicular magnetization component $M_z$ of the underlying (Ga,Fe)Sb, as presented in our previous work[S3]. However, we expect that the MPE occurs more strongly in the center 2D channel than in the edge 1D channels. This is because the $M_z$ component is smaller in the edges of (Ga,Fe)Sb where the magnetic moment of Fe usually is tilted towards the side surface. Therefore, we think that *the step-like increase of OMR at $I_C$ = 200 nA (Fig. 2a) is possibly caused by the sudden enhancement of the MPE in the edge due to expansion of the electron wavefunctions in the 1D edge channel towards the 2D center channel at this critical current value.*

As illustrated in Supplementary Fig. 5b, in the 1D edge channel, electron wavefunctions are confined by a triangular potential at the side surfaces and have limited penetration to the 2D center channel. When we increase $I$, however, the current is more concentrated in the edge, which has higher conductivity because of weaker magnetic scattering from MPE. This increases the electron carrier concentration in the edge. These changes may eventually lead to the occupation of the next quantized level at a slightly higher energy, whose electron wavefunction overlaps more largely with the 2D channel due to the weaker confinement. This enhances the 1D (edge) - 2D (center) wavefunction overlapping and consequently increases the MPE in the edge channel in a sudden manner as observed at $I_C$ = 200 nA, leading to the sudden increase in $\Delta R/R_0$, as shown in Fig. 2a. With more magnetic scattering in the edge transport, this also explains the slight increase of the total resistance $R_0$ at $I_C$ in Fig. 2a.

## Supplementary Note 3: Hall-bar-orientation dependence of OMR

The Hall-bar orientation (current direction) presented in this paper is always along the $[\bar{1}10]$ axis of the GaAs substrate. In order to investigate the effect of crystal orientation, we fabricate a Hall bar device aligned along the [110] direction and compare its magnetotransport data with those of the Hall bar device aligned along the $[\bar{1}10]$ direction (The device aligned along $[\bar{1}10]$ is the same as $D_L$). Supplementary Fig. 6a shows temperature dependence of the four-terminal resistance $R_{23}$ in the two devices where the current $I$ is applied parallel to [110] and $[\bar{1}10]$, respectively. The resistance at each $T$ differs between the two Hall bars with different orientations. Also, the magnetoresistance measurements exhibit different OMR magnitude as shown in Supplementary Fig. 6b. The OMR magnitude (= $[R_{23}(1\text{ T}) - R_{23}(-1\text{ T})]/2R_{23}(0\text{ T})$) of [110] and $[\bar{1}10]$ are 0.065% and 0.036%, respectively.

According to the previous study of the Rashba and Dresselhaus effects of InAs/GaSb[S4], the Dresselhaus effect is relatively small, less than one-sixth of the Rashba effect. Thus, it is unlikely that the differences in $R_{23}$ and the OMR magnitude originate from the Dresselhaus effect. One possible reason is the anisotropic distribution of the Fe atoms in the (Ga,Fe)Sb layer with an Fe content of ~20% along the two directions, [110] and $[\bar{1}10]$. In heavily Fe-doped (Ga,Fe)Sb (Fe > ~20%) such as that in our InAs/(Ga,Fe)Sb samples, it is known that spinodal decomposition occurs, leading to fluctuation in the local Fe concentration in the host GaSb crystal[S5]. This Fe-rich (Ga,Fe)Sb regions can favorably form in one direction, [110] or $[\bar{1}10]$[S6]. If this is the case, it can lead to different strength of the magnetic proximity effect (MPE) when the electron carriers in InAs flow in different directions, which will lead to anisotropic OMR.

## Supplementary Note 4: Equivalent circuit model of two- and four-terminal magnetotransport measurement

The two results shown in Fig. 2b and c can be understood by the equivalent circuit model shown in Supplementary Fig. 7. We describe the 4-terminal resistance as the sum of odd and even components. Here, the 4-terminal resistances facing each other ($R_{23}$ and $R_{65}$) are given by

$$R_{23}(B) = R_{23}^{\text{odd}}(B) + R_{23}^{\text{even}}(B) \qquad (S2)$$

and

$$R_{65}(B) = R_{65}^{\text{odd}}(B) + R_{65}^{\text{even}}(B) \qquad (S2')$$

Here, we assume that the even components are common in these two resistances

($R^{\text{even}}{}_{23}(B)$ = $R^{\text{even}}{}_{65}(B)$). Reflecting the edge transport data shown in Fig. 2b, the odd components $R_{23}{}^{\text{odd}}$ and $R_{65}{}^{\text{odd}}$ satisfy

$$R_{23}^{\text{odd}}(B) = -R_{65}^{\text{odd}}(B) \tag{S3}$$

Also, the 2D conduction does not show the odd component:

$$R_{2D}(B) = R_{2D}(-B) \tag{S4}$$

The 2-terminal resistance $R_{14}$ can be described as

$$\frac{1}{R_{14}(B)} = \frac{1}{R_{23}^{\text{odd}}(B) + R_{23}^{\text{even}}(B)} + \frac{1}{R_{65}^{\text{odd}}(B) + R_{65}^{\text{even}}(B)} + \frac{1}{R_{2D}(B)} \tag{S5}$$

Using eq. (S4) and (S5),

$$R_{14}(B) = \frac{R_{2D}(B)[(R_{23}^{\text{even}}(B))^2 - (R_{23}^{\text{odd}}(B))^2]}{2R_{23}^{\text{even}}(B)R_{2D}(B) + [(R_{23}^{\text{even}}(B))^2 - (R_{23}^{\text{odd}}(B))^2]} \tag{S6}$$

Therefore, $R_{14}$ is an even function of $B$.

## Supplementary Note 5: Counterevidence of intermixing from the Hall resistance

In order to check the Hall effect contribution as a possible origin of the OMR, we conducted two-terminal magnetotransport measurement, and obtained current and gate voltage dependence of the Hall effect. Although the four-terminal measurement can avoid the extrinsic resistance in the transport measurement, it may have the possibility of intermixing of the Hall resistance and the longitudinal resistance.

To confirm that the OMR is not caused by the Hall resistance, we measured the two-terminal resistance ($R_{23} = V_{23}/I_{23}$) in two InAs/(Ga,Fe)Sb devices as shown in Fig. 3a and b, where the dashed lines denote the outlines of the Hall bars. Device D1 in Fig. 3a is a Hall bar where the Au pads slightly touch on the edges, while device D2 in Fig. 3b is a Hall bar with branches in full contact with the Au pads. In D1, the two-terminal resistance $R_{23}$ contains large contact resistances, thus exhibiting a large parabolic MR as shown in the upper panel of Supplementary Fig. 8a. Nevertheless, an OMR is observed as shown in the bottom panel of Supplementary Fig. 8a. The small OMR is caused by the high contact resistances due to the small contact areas of the Au electrodes. On the other hand, in device D2, as shown in Supplementary Fig. 8 b, the OMR becomes dominant even at small magnetic fields of ±1 T, because the contact resistances are much lower. Therefore, these experiments show that the OMR effect appears not only in the four-terminal but also in the two-terminal configurations.

Finally, the current dependence of the Hall resistance ($=V_{26}/I_{14}$) supports this fact. As we mentioned in the main manuscript, the $R_0$ and $\Delta R/R_0$ show the step-like current dependence (Fig. 2a). On the other hand, the Hall resistance does not change with current as shown in Supplementary Fig. 9a. Also, $V_g$ dependence of the Hall resistance of D2 shows the different behaviour of OMR: As shown in Fig. 3d, the sign change of OMR is seen by the G1 gate voltage application, whereas the Hall resistance shows negative slope in every $V_g$ value (= +7, 0, −7 V) as shown in Supplementary Fig. 9b. These results strongly indicate that the OMR is *not* originated from the Hall effect.

## Supplementary Note 6: Comparison of the 1D and 2D transport via the device size effect

To estimate each contribution of the 1D edge and 2D layer transport channels, we performed transport measurements on two Hall bars, $D_L$ and $D_S$, with different sizes ($l_{14}$, $l_{23}$, $w$) = (600 μm, 200 μm, 100 μm) and (180 μm, 60 μm, 30 μm), respectively. Here $w$ is the width of the Hall bar and $l_{14}$, $l_{23}$ are the distances between terminals "1" to "4" and "2" to "3", respectively, as shown in Supplementary Fig. 11a. It is highly challenging to control the 1D and 2D conduction independently using the top gate voltage because the width of the 1D channel is too narrow. As described below, we compare the magnetotransport results and OMRs in the two Hall bars without applying a gate voltage.

As shown in Supplementary Fig. 11b, temperature ($T$) dependence of the four-terminal resistance $R_{23}$ shows significant difference between $D_L$ and $D_S$. Because the ratios $w$:$l_{23}$ of $D_L$ and $D_S$ are the same (=1:2), $R_{23}$ should be equal in $D_L$ and $D_S$ if the electrical conduction is uniform. However, the experimental $R_{23} - T$ curves differ between the two devices, which suggests that the electrical transport is non-uniform due to the coexistence of the 1D and 2D channels. As shown in the inset of Supplementary Fig. 11c, the transport results can be understood by a simple resistor network model, where the resistors corresponding to the 2D ($R_{2D}$) and 1D ($R_{1D}$) channels are connected in parallel. Assuming that the 1D channel has the same width in $D_L$ and $D_S$, the total resistance of the network can be expressed as:

$$(R_{23}(0\ \mathrm{T}))^{-1} = \left(r_{2D}\frac{l_{23}}{w}\right)^{-1} + (r_{1D}l_{23})^{-1} \tag{S7}$$

where $R_{2D} = r_{2D}(l_{23}/w)$, and $R_{1D} = r_{1D}l_{23}$. By solving simultaneous equations for $r_{2D}$ and $r_{1D}$ with $D_L$ and $D_S$ at each temperature, we obtain separate $R_{2D} - T$ and $R_{1D} - T$ curves as shown in Supplementary Fig. 11c. Note that Supplementary Fig. 11c shows the case of $D_L$. At 3.8 K, the ratio $R_{1D}/R_{2D}$ is 3.2, which corresponds to the current distribution ratio between the 1D and 2D channels. Thus, *2D and 1D transport channels coexist in the InAs/(Ga,Fe)Sb heterostructures, and the 1D transport does not dominate*.

One interesting observation is that the behavior of the $R_{2D} - T$ and $R_{1D} - T$ curves below 10 K agree with our model presented in Supplementary Note 2. In this Note, we argued that there is scattering between in the 1D and 2D channel, where the 2D channel has much stronger MPE than the 1D channel. As shown in Supplementary Fig. 11c, the 2D channel resistance exhibits an increase as temperature decreases below 10 K, which follows the logarithmic trend that is characteristic of the Kondo effect. This suggests strong scattering with magnetic impurities at the InAs/(Ga,Fe)Sb interface in the 2D channel. In contrast, the 1D channel resistance decreases as temperature decreases, exhibiting metallic conduction. This fact implies that the electrons in the 2D channel feel stronger MPE from the localized spins in the (Ga,Fe)Sb layer underneath, just as we expected.

The magnetotransport data of $R_{23}$ (four-terminal resistance) are shown in the Supplementary Fig. 11d. The OMR magnitudes at 1 T (=[$R_{23}$(1 T) − $R_{23}$(−1 T)]/2$R_{23}$(0 T)) of $D_L$ and $D_S$ are 0.037% and 0.095%, respectively. The OMR decreases with increasing the Hall bar width $w$, which is reasonable considering our scenario of parallel conduction between the 1D and 2D channels: When $w$ increases, the conduction of the 2D channel, which does not show OMR, becomes more dominant. Thus, the ratio between the resistance change due to OMR in the 1D edge channel versus the total resistance becomes smaller, leading to a smaller OMR.

This can be proved analytically as described below. Using eq. (S7), the OMR magnitude is expressed as

$$
\frac{R_{23}(1\ \mathrm{T}) - R_{23}(-1\ \mathrm{T})}{2R_{23}(0\ \mathrm{T})}
$$
$$
= \frac{\left[\left(R_{\mathrm{2D}}(1\ \mathrm{T})\right)^{-1} + \left(R_{\mathrm{1D}}(1\ \mathrm{T})\right)^{-1}\right]^{-1} - \left[\left(R_{\mathrm{2D}}(-1\ \mathrm{T})\right)^{-1} + \left(R_{\mathrm{1D}}(-1\ \mathrm{T})\right)^{-1}\right]^{-1}}{2\left[\left(R_{\mathrm{2D}}(0\ \mathrm{T})\right)^{-1} + \left(R_{\mathrm{1D}}(0\ \mathrm{T})\right)^{-1}\right]^{-1}}
$$
$$
= \frac{\left[\left(wr_{\mathrm{2D}}(1\ \mathrm{T})\right)^{-1} + \left(r_{\mathrm{1D}}(1\ \mathrm{T})\right)^{-1}\right]^{-1} - \left[\left(wr_{\mathrm{2D}}(-1\ \mathrm{T})\right)^{-1} + \left(r_{\mathrm{1D}}(-1\ \mathrm{T})\right)^{-1}\right]^{-1}}{2\left[\left(wr_{\mathrm{2D}}(0\ \mathrm{T})\right)^{-1} + \left(r_{\mathrm{1D}}(0\ \mathrm{T})\right)^{-1}\right]^{-1}} \tag{S8}
$$

The differential of eq. (S8) with *w* is

$$
\frac{\partial}{\partial w}\left(\frac{R_{23}(1\ \mathrm{T}) - R_{23}(-1\ \mathrm{T})}{2R_{23}(0\ \mathrm{T})}\right)
$$
$$
= \frac{\partial}{\partial w}\left(\frac{[(h'w)^{-1} + g_{+}^{-1}]^{-1} - [(h'w)^{-1} + g_{-}^{-1}]^{-1}}{2[(h_0 w)^{-1} + g_0^{-1}]^{-1}}\right)
$$
$$
= -\frac{(g_{-} - g_{+})\left(h_0(h'^2 w^2 - g_{+}g_{-}) + h' g_0(2h'w + g_{+} + g_{-})\right)}{2(h'w + g_{+})^2 (h'w + g_{-})^2} \tag{S9}
$$

where $1/r_{\mathrm{1D}}(\pm 1\ \mathrm{T}) = g_{\pm}, 1/r_{\mathrm{1D}}(0\ \mathrm{T}) = g_0, 1/r_{\mathrm{2D}}(\pm 1\ \mathrm{T}) = h',$ and $1/r_{\mathrm{2D}}(0\ \mathrm{T}) = h_0$. Note that since the 2D channel does not show the OMR component, $R_{\mathrm{2D}}(1\ \mathrm{T}) = R_{\mathrm{2D}}(-1\ \mathrm{T}),$ *i.e.* $r_{\mathrm{2D}}(1\ \mathrm{T}) = r_{\mathrm{2D}}(-1\ \mathrm{T}) = 1/h'$.

Here we prove that eq. (S9) is always negative at $w > 0$. For the first bracket on the numerator, since the OMR magnitude in eq. (S8) is defined as positive, $R_{\mathrm{1D}}(1\ \mathrm{T}) > R_{\mathrm{1D}}(-1\ \mathrm{T})$, *i.e.* $r_{\mathrm{1D}}(1\ \mathrm{T}) > r_{\mathrm{1D}}(-1\ \mathrm{T})$. Therefore, $g_{-} - g_{+} > 0$. Also, for the second bracket on the numerator, since all the parameters are positive, it is sufficient to prove $h'^2 w^2 - g_{-}g_{+} > 0$. While $R_{\mathrm{1D}}(0\ \mathrm{T})/R_{\mathrm{2D}}(0\ \mathrm{T}) = 3.2$, the MR magnitude at ±1 T is less than 6% as shown in Supplementary Fig. 11d, which implies $R_{\mathrm{2D}}(\pm 1\ \mathrm{T}) < R_{\mathrm{1D}}(\pm 1\ \mathrm{T})$. Thus,

$$
R_{\mathrm{2D}}(1\ \mathrm{T})R_{\mathrm{2D}}(-1\ \mathrm{T}) < R_{\mathrm{1D}}(1\ \mathrm{T})R_{\mathrm{1D}}(-1\ \mathrm{T})
$$
$$
\frac{\left(r_{2D}(1\ \mathrm{T})\right)^2}{w^2} < r_{1D}(1\ \mathrm{T})r_{1D}(-1\ \mathrm{T})
$$
$$
h'^2 w^2 - g_{-}g_{+} > 0 \tag{S10}
$$

From the argument described above, the OMR magnitude decreases with increasing *w*. Thus, our model can explain the device size dependence of OMR, indicating that the 1D channel is the main origin of the OMR.

## Supplementary Note 7: Theoretical analysis using Boltzmann's equation

The low-energy 1D electrons with a Rashba-type SOI at the edge of a 2D electron gas on a ferromagnetic insulator (FI) are described by the effective Hamiltonian given by eq. (1) in the main manuscript,

$$
\widehat{H}_{\mathrm{1D}}(k_x) = \frac{\hbar^2 k_x^2}{2m^*}\sigma_0 + (\varLambda_{\mathrm{side}}k_x + \Delta_z)\sigma_z + \varLambda_{\mathrm{top}}k_x\sigma_y \tag{1}
$$

The 2D electrons near the interface are coupled to the magnetic dopants (Fe) in (Ga,Fe)Sb via the *s-d* exchange interaction, described by

$$\hat{H}_{sd}(\mathbf{r}) = -V_{sd}\sum_i \mathbf{S}_i \cdot \boldsymbol{\sigma}\delta(\mathbf{r}-\mathbf{R}_i) \tag{S11}$$

where $V_{\mathrm{sd}}$ is the *s-d* exchange potential, $\mathbf{S}_i$ is the local spin operator, and $\mathbf{R}_i$ is the position operator of the *i*th Fe magnetic dopant. Equation (1) gives the energy dispersion shown in eq. (2),

$$E_s = \frac{\hbar^2 k_x^2}{2m^*} + s\sqrt{(\Lambda_{\mathrm{side}}k_x + \Delta_z)^2 + \left(\Lambda_{\mathrm{top}}k_x\right)^2} \tag{2}$$

where $s = +/-$ denotes the upper and lower bands, as well as the related eigenstates $\varphi_{ks}(x) = e^{ikx}|u_{ks}\rangle$, with

$$|u_{k+}\rangle = \begin{pmatrix} -i\sin\frac{\theta_k}{2} \\ \cos\frac{\theta_k}{2} \end{pmatrix}, |u_{k-}\rangle = \begin{pmatrix} -i\cos\frac{\theta_k}{2} \\ \sin\frac{\theta_k}{2} \end{pmatrix} \tag{S12}$$

Here, the angle $\theta_k$ is defined by

$$\cos\theta_k = \frac{\Lambda_{\mathrm{side}}k_x + \Delta_z}{\sqrt{(\Lambda_{\mathrm{side}}k_x + \Delta_z)^2 + \left(\Lambda_{\mathrm{top}}k_x\right)^2}}, \sin\theta_k = \frac{\Lambda_{\mathrm{top}}k_x}{\sqrt{(\Lambda_{\mathrm{side}}k_x + \Delta_z)^2 + \left(\Lambda_{\mathrm{top}}k_x\right)^2}} \tag{S13}$$

The band diagrams are schematically represented in Fig. 4a, in which the position of the Fermi energy determines the topology of the Fermi surfaces. In the presence of an out-of-plane magnetic field ($B_z$), the Rashba-type spin splitting becomes asymmetric because of the Zeeman splitting $\Delta_z$. As we mentioned in the main manuscript, due to $\Lambda_{\mathrm{top}}$, the eigenstate of eq. (2) can be labelled by chirality, which is indicated by green and pink colors in Fig. 4a and b.

Let us calculate the charge current arising from the edge transport. According to eq. (1), the velocity operator is given by

$$\hat{v} = \frac{1}{\hbar}\frac{\partial \hat{H}_{1D}}{\partial k_x} = \frac{\hbar k_x}{m^*}\sigma_0 - \frac{\Lambda_{\mathrm{side}}}{\hbar}\sigma_z + \frac{\Lambda_{\mathrm{top}}}{\hbar}\sigma_y \tag{S14}$$

The expectation value of Eq. (S14) on each eigenstate corresponds to the electron group velocity

$$v_s^{(0)} = \frac{1}{\hbar}\frac{\partial E_s}{\partial k_x} = \langle u_{ks}|\hat{v}|u_{ks}\rangle = \frac{\hbar k_x}{m^*} + s\frac{\Lambda_{\mathrm{side}}}{\hbar}\cos\theta_k + s\frac{\Lambda_{\mathrm{top}}}{\hbar}\sin\theta_k \tag{S15}$$

where the first term is the normal velocity and the second and third terms are additional velocities induced by the Rashba SOI. Hereafter, we assume $\Lambda_{\mathrm{top}}\,(=\hbar\lambda_{\mathrm{top}}) \ll \Lambda_{\mathrm{side}}\,(=\hbar\lambda_{\mathrm{side}})$, which means that the electric field at the side edges is much larger than that at the top surface, and neglect the effect of the Rashba SOI from the interface on the energy dispersion.[S7]

Let us now calculate the charge current driven by an electric field $E_x$. When the electric field is applied, under the relaxation time approximation, the Fermi surface shifts by $\delta k_x = -eE_x\tau_s/\hbar$, where $\tau_s$ is the electron relaxation time. $\tau_+$ and $\tau_-$ can differ due to the chirality dependent scattering. (See Supplementary Note 8.)

By taking a power series expansion with respect to the electric field $E_x$ up to the 1st order, the corresponding deviation from the equilibrium distribution function $f_s^{(0)}$ is given by

$$f_s = f_s^{(0)} + f_s^{(1)}(E_x), \tag{S16}$$

where $f_s^{(1)}$ is the first-order deviation from $f_s^{(0)}$. Therefore, the charge current density $J_x$

consists of $J_x^{(0)}$ and $J_x^{(1)}$; $J_x = J_x^{(0)} + J_x^{(1)}(E_x)$. Here, the first-order current density is given by

$$
\begin{aligned}
J_x^{(1)} &= \sum_s \int \frac{dk_x}{2\pi}\left(-ev_s^{(0)}\right) f_s^{(1)}(E_x) \\
&= -\frac{e^2 E_x}{2\pi\hbar}\sum_s \tau_s \int dk_x v_s^{(0)} \frac{\partial f_s^{(0)}}{\partial k_x} \\
&= C_1 \sum_s \tau_s \int dE_s v_s^{(0)}(E_s)\left(-\frac{\partial f_s^{(0)}}{\partial E_s}\right) \qquad \text{(S17)}
\end{aligned}
$$

where $C_1 = e^2E_x/2\pi\hbar$. For $T \rightarrow 0$, $-\partial f_s^{(0)}/\partial E_s = \delta(E_s - E_F)$; then,

$$
\begin{aligned}
J_x &= C_1 \sum_s \tau_s v_s^{(0)}(E_F) \\
&= C_1 \sum_s \tau_s |\lambda_E| \left(1 + \frac{2E_F}{m^*\lambda_{\text{side}}^2} - s\frac{|\lambda_E|}{\lambda_E}\frac{2\Delta_z}{m^*\lambda_{\text{side}}^2}\right)^{\frac{1}{2}} \\
&= C_1 \tau_- |\lambda_{\text{side}}| \left[\left(1 + \frac{2E_F}{m^*\lambda_{\text{side}}^2} + \frac{|\lambda_{\text{side}}|}{\lambda_{\text{side}}}\frac{2\Delta_z}{m^*\lambda_{\text{side}}^2}\right)^{\frac{1}{2}} + \alpha\left(1 + \frac{2E_F}{m^*\lambda_{\text{side}}^2} - \frac{|\lambda_{\text{side}}|}{\lambda_{\text{side}}}\frac{2\Delta_z}{m^*\lambda_{\text{side}}^2}\right)^{\frac{1}{2}}\right] \qquad \text{(S18)}
\end{aligned}
$$

For $E_F \gg \Delta_z$, eq. (S18) is approximately rewritten as

$$
J_x \simeq C_1 \tau_- |\lambda_{\text{side}}| \sqrt{1 + \frac{2E_F}{m^*\lambda_{\text{side}}^2}}\left[1 + \alpha + (1-\alpha)\frac{|\lambda_{\text{side}}|}{\lambda_{\text{side}}}\frac{\Delta_z}{2E_F + m^*\lambda_{\text{side}}^2}\right] \qquad \text{(S19)}
$$

where

$$
\alpha = \frac{\tau_+}{\tau_-} \qquad \text{(S20)}
$$

is a parameter, which results from an asymmetric scattering between the $E_-$ and $E_+$ bands. The detailed discussion on this parameter is given in Supplementary Note 8. Therefore, the conductivity is given by eq. (5),

$$
\sigma_{xx} \simeq \frac{e^2}{h}\tau_- |\lambda_{\text{side}}| \sqrt{1 + \frac{2E_F}{m^*\lambda_{\text{side}}^2}}\left[1 + \alpha + (1-\alpha)\frac{|\lambda_{\text{side}}|}{\lambda_{\text{side}}}\frac{g\mu_B}{2E_F + m^*\lambda_{\text{side}}^2}B_z\right] \qquad (5)
$$

where $h$ is the Planck's constant.

## Supplementary Note 8: Origin of the asymmetric scattering

While the influence of ferromagnetism on the OMR is common in all the previous reports and ours, there is one fundamental difference between the OMR observed in our InAs/(Ga,Fe)Sb bilayer heterostructure and the others. As shown in Supplementary Table 1, in the previous works, the linear OMR only occurred when the magnetic field **B** and the magnetization **M** are *separately* changed (thus they are not always parallel; **B** $\nparallel$ **M**). This is because, although each of **B** and **M** can break the TRS, simultaneous reversal of both **B** and **M** preserves the TRS. Therefore, when **B** and **M** are completely aligned (**B** // **M**, which is the case when **B** is large), the Onsager's reciprocal theorem requires $\sigma_{xx}(\mathbf{B},\mathbf{M}) = \sigma_{xx}(-\mathbf{B},-\mathbf{M})$ and no OMR is allowed ($\sigma_{xx}$ is longitudinal conductivity), as discussed in Ref. S10. However, when $\mathbf{B} \nparallel \mathbf{M}$, the TRS is broken by reversing **B** or **M** alone, which relaxes the Onsager's theorem requirement and allows OMR to occur. For example, the OMR in $SmCo_5$[S8,S9] is observed only when **B** is smaller than the coercivity of **M** (~ 2 T). Thus, previously reported OMR phenomena were only realized in small magnetic field regions.

In contrast, the OMR in our InAs/(Ga,Fe)Sb system is fundamentally different. In our case, the OMR is present and linearly proportional to the magnetic field **B** in the whole range of **B** up to 10 T, which is much larger than the coercivity of (Ga,Fe)Sb (~ 50 mT). It is obvious that the magnetization **M** of (Ga,Fe)Sb should closely follow **B** in most of the magnetic field range (*i.e.* **B** // **M**). Therefore, the observation of OMR in our case is striking, because the TRS is preserved when both **B** and **M** are simultaneously reversed and always parallel, as mentioned above. Thus, this suggests that the OMR in our InAs/(Ga,Fe)Sb system should have a different origin and cannot be explained by the same theoretical framework of the previous reports.

At the present stage, we do not have a rigorous theoretical explanation for the large OMR observed in our InAs/(Ga,Fe)Sb system. However, our idea is that there should be some other factors that break the TRS, which only weakly depends on the external magnetic field **B**. Here we show a possible mechanism to explain our OMR results. Figure 4a in the main manuscript illustrates the band dispersions of the 1D edge channel of InAs, where there are two branches of energy dispersion $E_+$ and $E_-$. These dispersions are the results of the Rashba spin orbit interaction (SOI) at the top ($z$ direction) and side ($y$ direction) surface of the InAs edge (see Fig.4a in the main manuscript). The eigenvalues of these $E_+$ and $E_-$ branches were obtained from the Hamiltonian in eq. (1) and have been given in the main manuscript.

$$\hat{H}_{1\mathrm{D}}(k_x) = \frac{\hbar^2 k_x^2}{2m^*}\sigma_0 + (\Lambda_{\mathrm{side}} k_x + \Delta_z)\sigma_z + \Lambda_{\mathrm{top}} k_x \sigma_y \qquad (1)$$

It is important to realize that the spin components $\sigma_y$ and $\sigma_z$ of the electron carriers are locked to the momentum $k_x$ in opposite directions between $E_+$ and $E_-$. Thus, the + and – subscripts also correspond to the different "chirality" of these bands. In our theoretical model based on the Boltzmann formalism, we proposed a phenomenological parameter, $\alpha$ (= $\tau_+/\tau_-$, where $\tau_+$ and $\tau_-$ are the relaxation time of electron carriers in the $E_+$ and $E_-$ states, respectively). If there is asymmetry between $\tau_+$ and $\tau_-$ (that is, $\alpha \neq 1$), the OMR is expressed as:

$$\sigma_{xx} \simeq \frac{e^2}{h}\,\tau_-|\lambda_{\text{side}}|\sqrt{1+\frac{2E_F}{m^*\lambda_{\text{side}}^2}}\left[(1-\alpha)\frac{|\lambda_{\text{side}}|}{\lambda_{\text{side}}}\frac{g\mu_B}{2E_F+m^*\lambda_{\text{side}}^2}B_z\right] \qquad (5)$$

This equation can quantitatively reproduce the linear OMR results observed in our experiment when $\alpha \neq 1$, as shown in Fig. 4c and d in the main manuscript. One possible origin of the asymmetric relaxation time between $E_+$ and $E_-$ can be deduced if we consider the scattering from the 1D edge channel to the 2D channel in the InAs layer, as shown in Supplementary Fig. 5a. In the 2D channel, the spin component $\sigma_y$ of the electron carriers is also locked to $k_x$ due to the Rashba effect due to the electric field in the $z$ direction. Here we consider that only the scattering within the same chirality is allowed (Supplementary Fig. 15a). The relaxation time of each $\sigma_y$ direction ($\tau_+$, $\tau_-$) should be different between the + and – chirality due to their different density of states at the Fermi level. Because the chirality + and – are only determined by the Rashba SOI, the definition and magnitude of $\alpha$ do not change even when reversing the $z$ component of **B** (= $B_z$). This leads to the appearance of OMR even when **B**//**M** (Supplementary Fig. 15b). We note that the $\sigma_z$ component of electron carriers may not be conserved in the scattering between the 1D and 2D channels because spin-flip scattering events may occur with the existence of localized spins in (Ga,Fe)Sb, which are aligned in the $z$ direction. The current-independence of the OMR in our system can also be explained using the same framework: When reversing the current direction, $\sigma_y$ in both 1D and 2D channels are flipped, and the definition of $\alpha$ does not change (Supplementary Fig. 15c). Thus, the TRS is broken when focusing only on the 1D channel while the non-reciprocity does not occur when considering both the 1D and 2D channels.

The magnetic proximity effect (MPE) also plays an important role in the scattering process. As we mention in Supplementary Note 2 and 6, the MPE mainly affects the 2D channel by opening a gap (= $\Delta_{2D}$) between different chirality bands as shown in Supplementary Fig. 15a and d; as the MPE is increased, $\Delta_{2D}$ is increased. As a result, the energy dispersion is altered by the MPE, which will lead to larger imbalance between $\tau_+$ and $\tau_-$ and larger OMR (Supplementary Fig. 15d).

**Supplementary Table 1. Comparison of OMR observed in previous reports and our work**. The maximum OMR magnitude $\Delta R/R_0$ (= $[(R(B)-R(-B))/2]/R(0\ \mathrm{T})$) normalized by $R_0$ (= $R(0\ \mathrm{T})$) were obtained under magnetic field $B$ at temperature $T$.

| **material** | **$\Delta R/R_0$ (%)** | **$B$ (T)** | **$T$ (K)** | **Observable under B // M** | **Proposed origin** | **ref.** |
|---|---|---|---|---|---|---|
| $SmCo_5$ | $1.3\times10^{-2}$ | 0.015 | room temp. | No | non-uniform distribution of the magnetization | S8 |
| $SmCo_5$ | $4.6\times10^{-2}$ | 0.5 | 300 | No | Zeeman splitting/ anomalous Hall effect | S9,S10 |
| $Gd_2Os_2O_7$ | $5.0\times10^{-2}$ | 2 | 195 | No | magnetic domain walls | S9 |
| $Eu_2Ir_2O_7$ (theory) | - | - | - | No | Berry curvature, magnetic moment, and shift vector | S10 |
| $Eu_2Ir_2O_7$ (experiment) | 0.44 | 9 | 2 | No | magnetic texture | S11 |
| $Fe_3GeTe_2$ /graphite/ $Fe_3GeTe_2$ | 1.1 | 0.01 | 50 | No | interfacial SOI of $Fe_3GeTe_2$ as a topological nodal line | S12 |
| **InAs/ (Ga,Fe)Sb** | **13.5**<br>**5** | **10**<br>**10** | **2**<br>**300** | **Yes** | **Rashba SOI at the edge of InAs and magnetic proximity effect** | **Our work** |

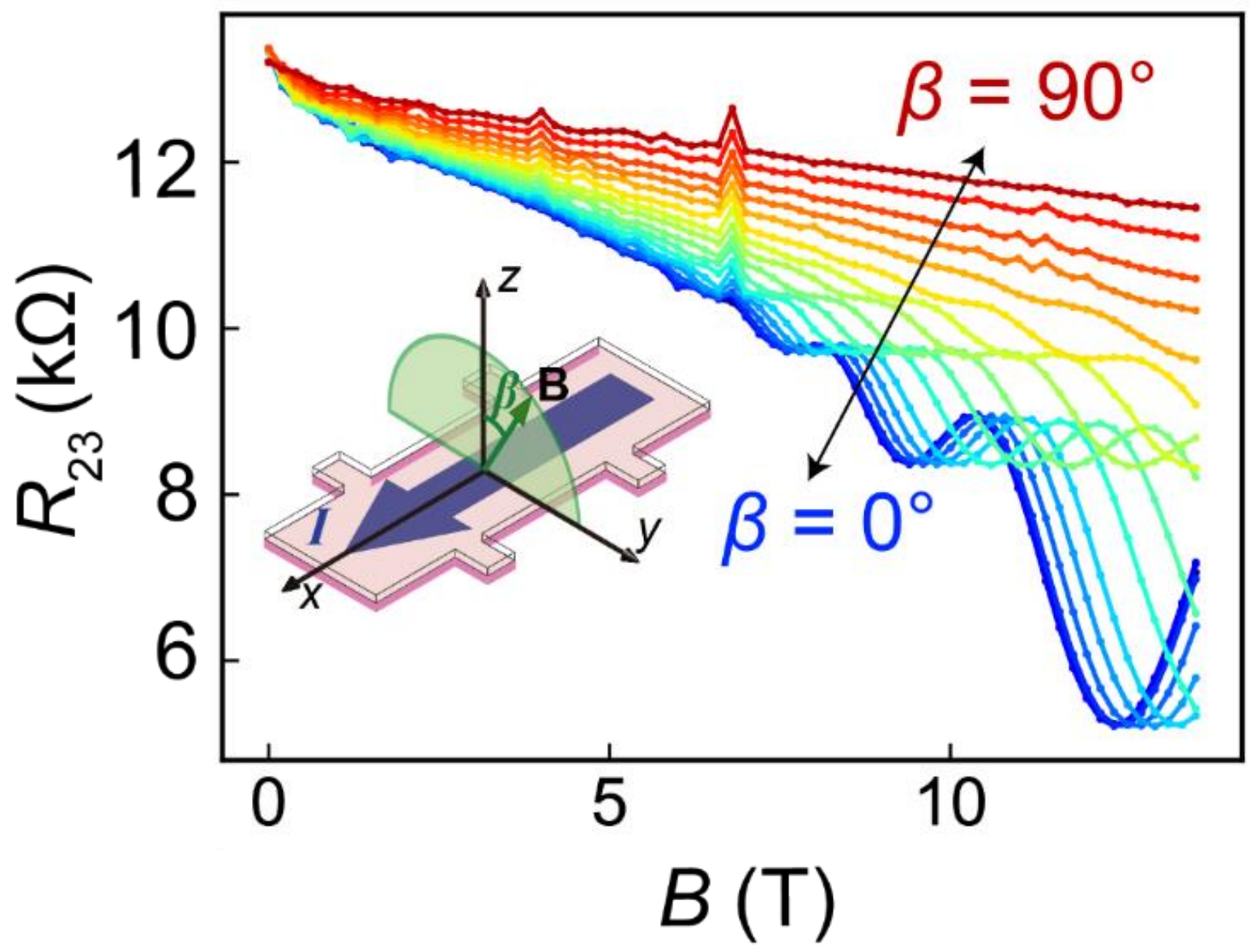

**Supplementary Fig. 1| Angular dependence of SdH oscillation.** Magnetoresistance of sample A in each $\beta$ angle from 0° to 90°. The inset shows the definition of $\beta$ in the *yz* plane.

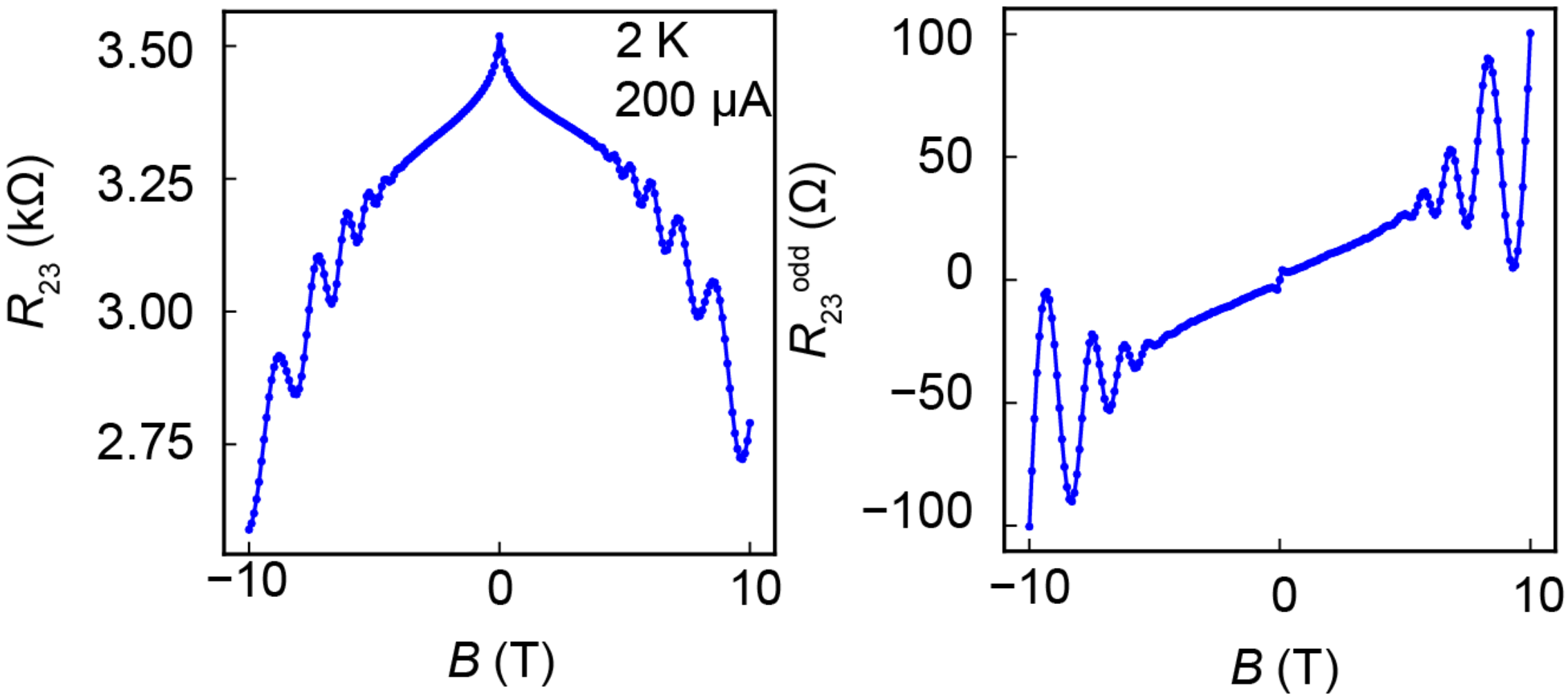

**Supplementary Fig. 2| Oscillating components in the OMR of a high mobility sample.** Magnetotransport measurement results of a higher mobility sample of the same structure shown in Fig. 1a (left panel) and the odd component (right panel) at 2 K with 200 μA.

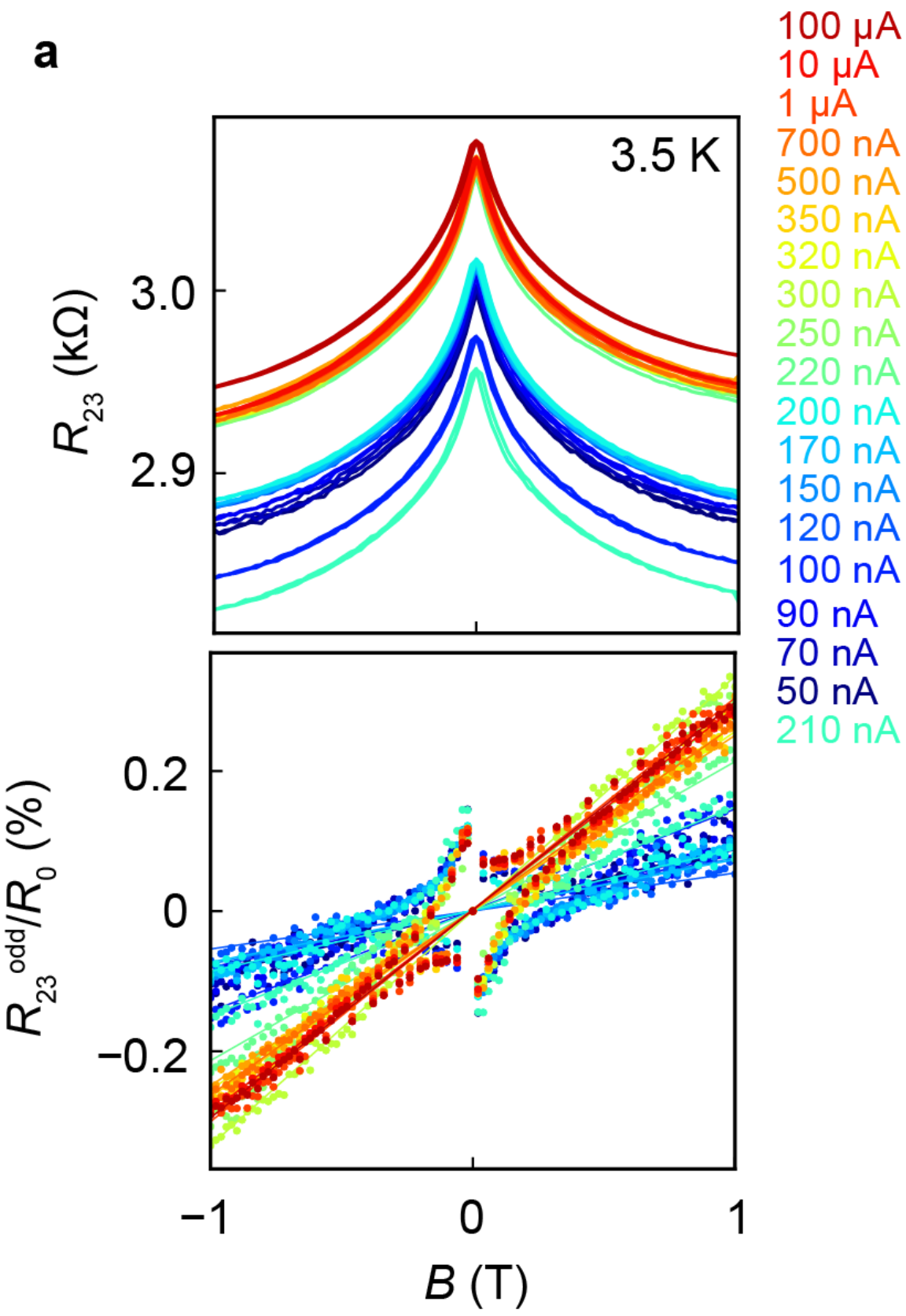

**Supplementary Fig. 3| Current dependence of the OMR**. **a**, Current dependence of the magnetoresistance curves (upper panel) and the odd components (lower panel) at 3.5 K with perpendicular $B$. Note that $R_{23}^{\mathrm{odd}}(B)= (R_{23}(B) - R_{23}(-B))/2$. Here, $R_{23}(B)$ is the resistance measured between terminals 2 and 3, and $R_0$ is $R_{23}(0\ \mathrm{T})$.

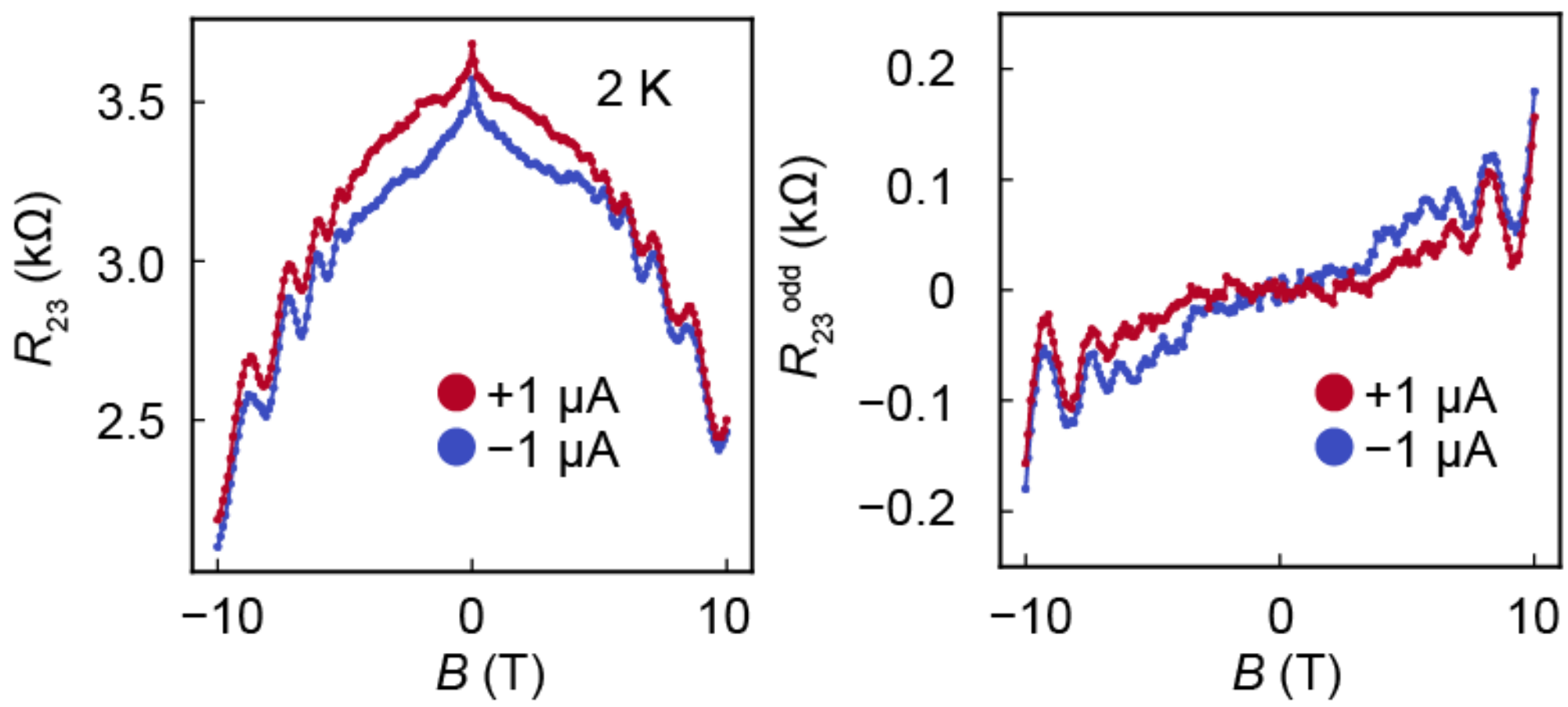

**Supplementary Fig. 4| Current direction dependence of the OMR.** Magnetoresistance curves with opposite current directions (left panel) and and their odd components (right panel) at 2 K with ±1 μA.

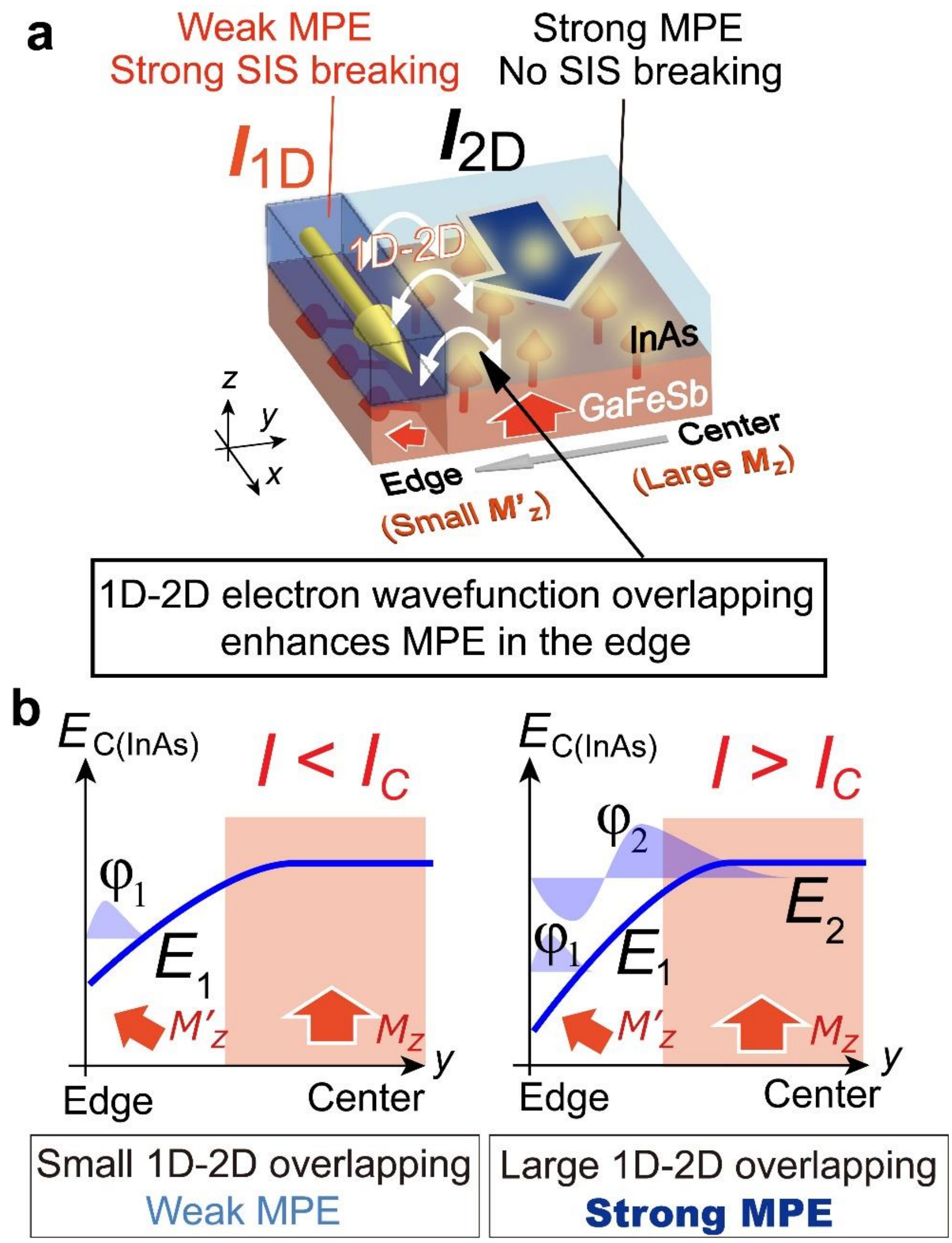

**Supplementary Fig. 5| Electronic states and magnetization situation near the side edge a,** Schematic illustration of the InAs/(Ga,Fe)Sb bilayer heterostructure near the Hall bar edge. In the (Ga,Fe)Sb layer, the magnetic moment $M_z$' (red arrows) near the side edge may be canted and does not effectively induce the MPE in the InAs edge channel. On the other hand, in the 2D channel (center) side, the magnetic moment $M_z$ is aligned in the $z$ direction due to the magnetic anisotropy of (Ga,Fe)Sb, leading to strong MPE. Through overlapping of the electron wavefunctions in the 1D channel (edge) with the 2D channel (center), MPE is strongly induced in the 1D channel at $I > I_C$. The MPE, together with the Rashba SOI, leads to the appearance of OMR. **b**, Illustrated electronic subband structure of the conduction band bottom $E_{C(InAs)}$ of InAs near the edge (blue curves). When the current $I$ is increased, the electron carriers are accumulated near the edge. This leads to a change in the occupied quantized levels at $I_C$. When higher levels are occupied by electrons at $I > I_C$, the 1D wavefunctions largely overlap with the 2D center region, which suddenly enhances the MPE in the edge, leading to the sudden increase of OMR.

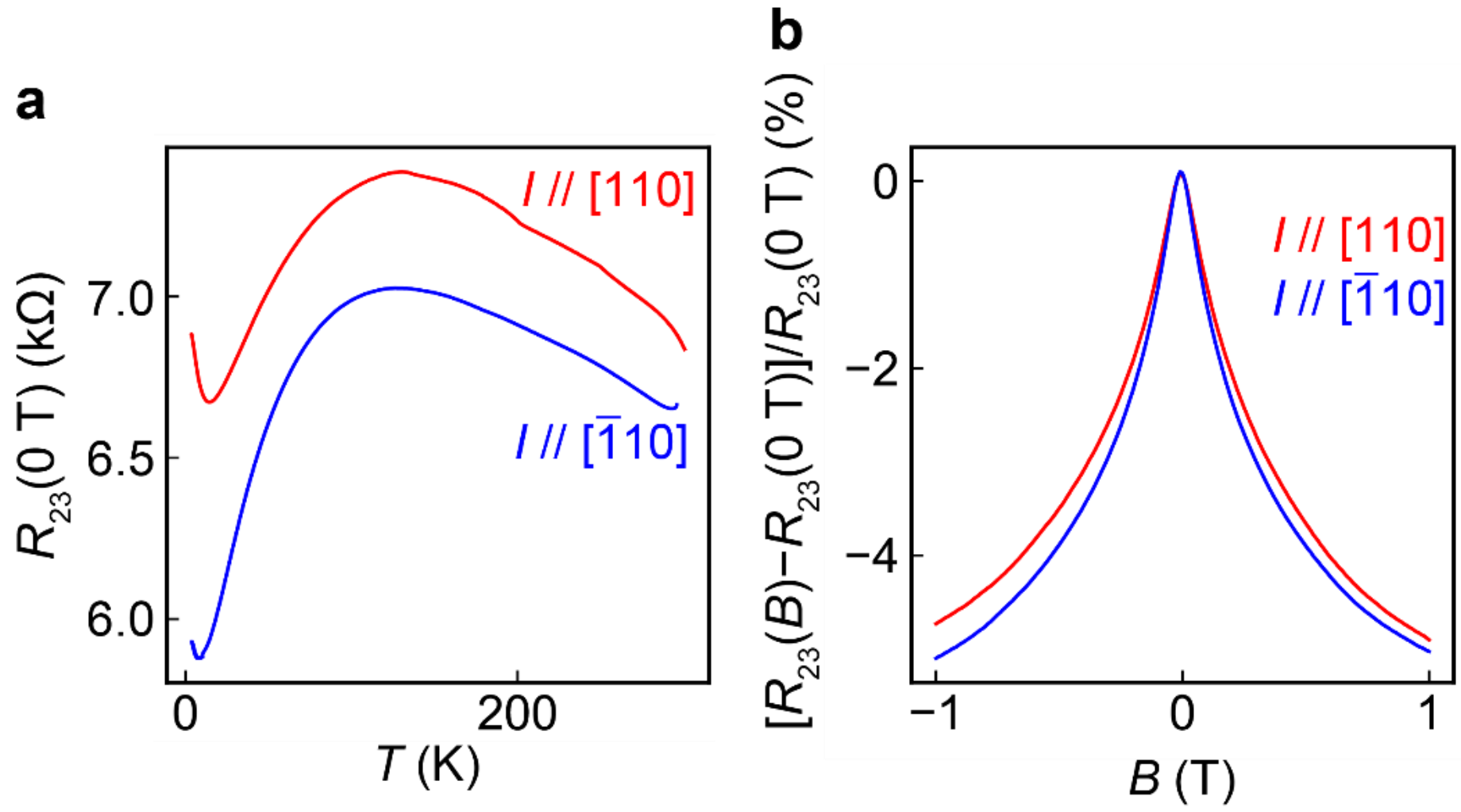

**Supplementary Fig. 6| Hall-bar-orientation dependence of electrical transport and OMR. a**, Temperature ($T$) dependence of $R_{23}$(0T) with $I$ // [110] (red) and $I$ // [$\bar{1}$10] (blue). **b**, Normalized magnetoresistance by the zero-field resistance, [$R_{23}$($B$)−$R_{23}$(0 T)]/ $R_{23}$(0 T), with $I$ // [110] (red) and $I$ // [$\bar{1}$10] (blue).

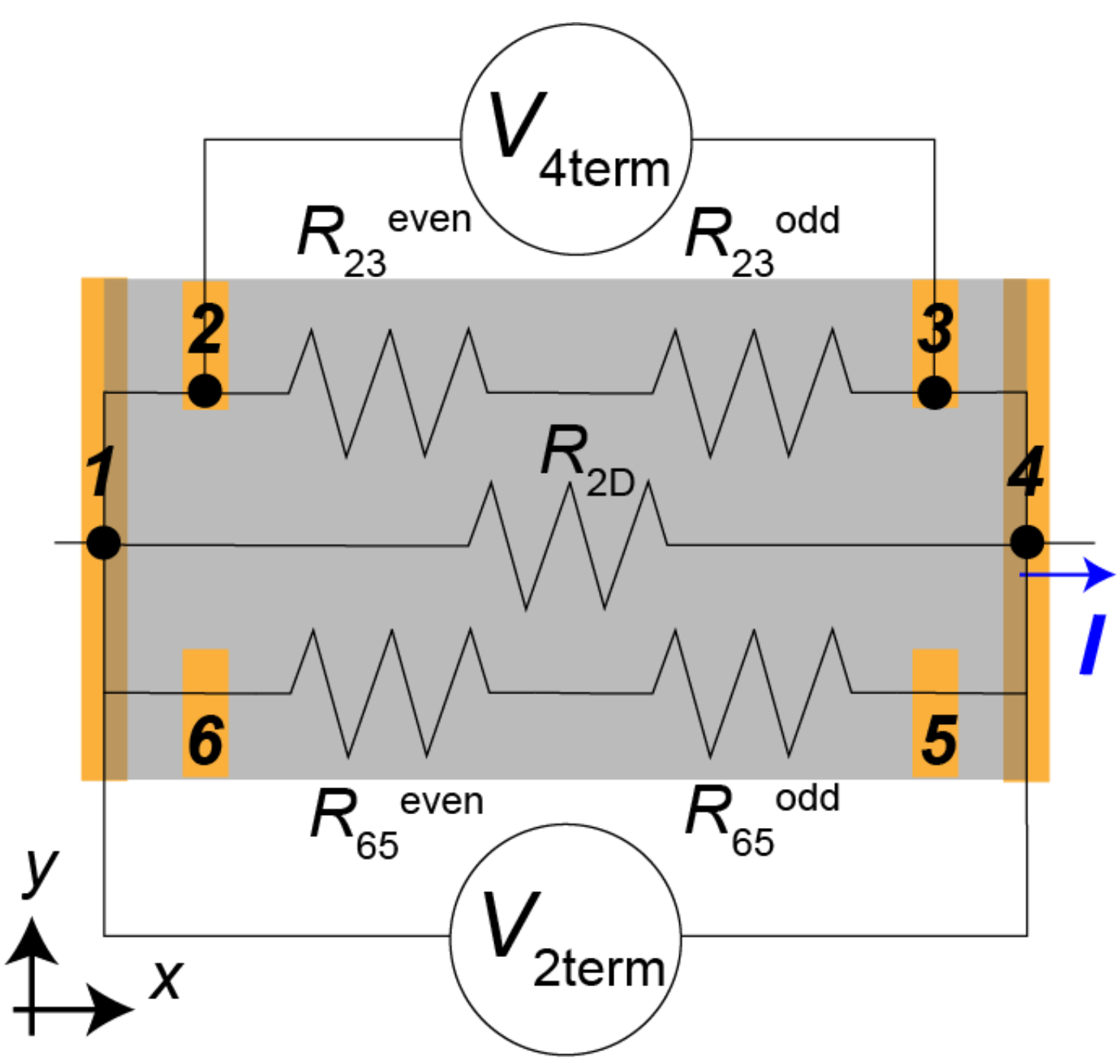

**Supplementary Fig. 7| Equivalent circuit model for the OMR in two- and four-terminal measurements.** Resistor network diagram representing our InAs/(Ga,Fe)Sb device and schematic diagram of the top view of our Hall bar device. $R_{23(65)}^{\mathrm{even}}$ and $R_{23(65)}^{\mathrm{odd}}$ represent the resistance components that are even and odd functions of the external magnetic field, respectively, observed in the upper (lower) terminals.

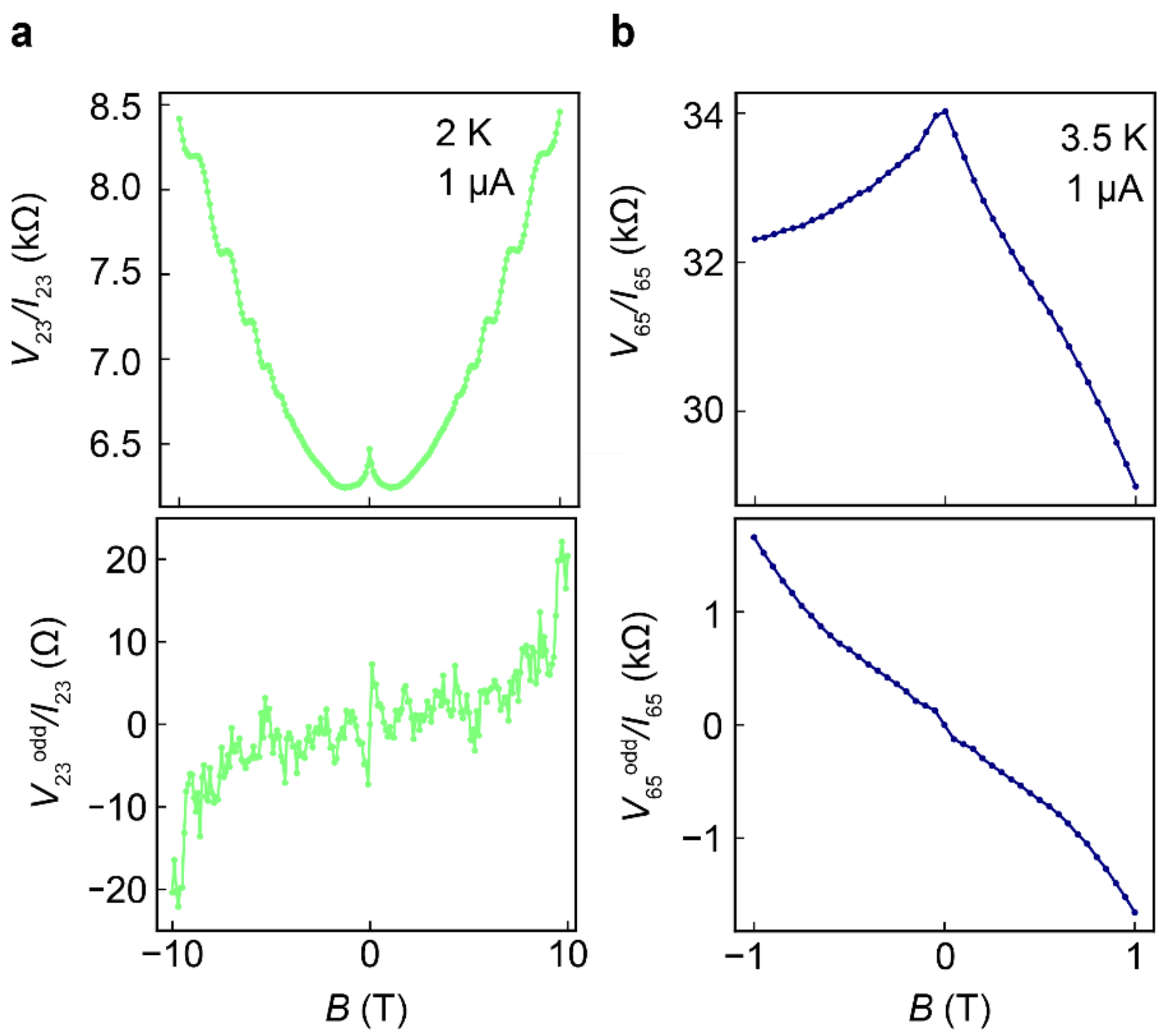

**Supplementary Fig. 8| Two-terminal measurement of D1 and D2. a**, Two-terminal measurement ($V_{23}/I_{23}$) of device D1 at 2 K with 1μA. **b**, Two-terminal measurement ($V_{65}/I_{65}$) of device D2 at 3.5 K with 1μA.

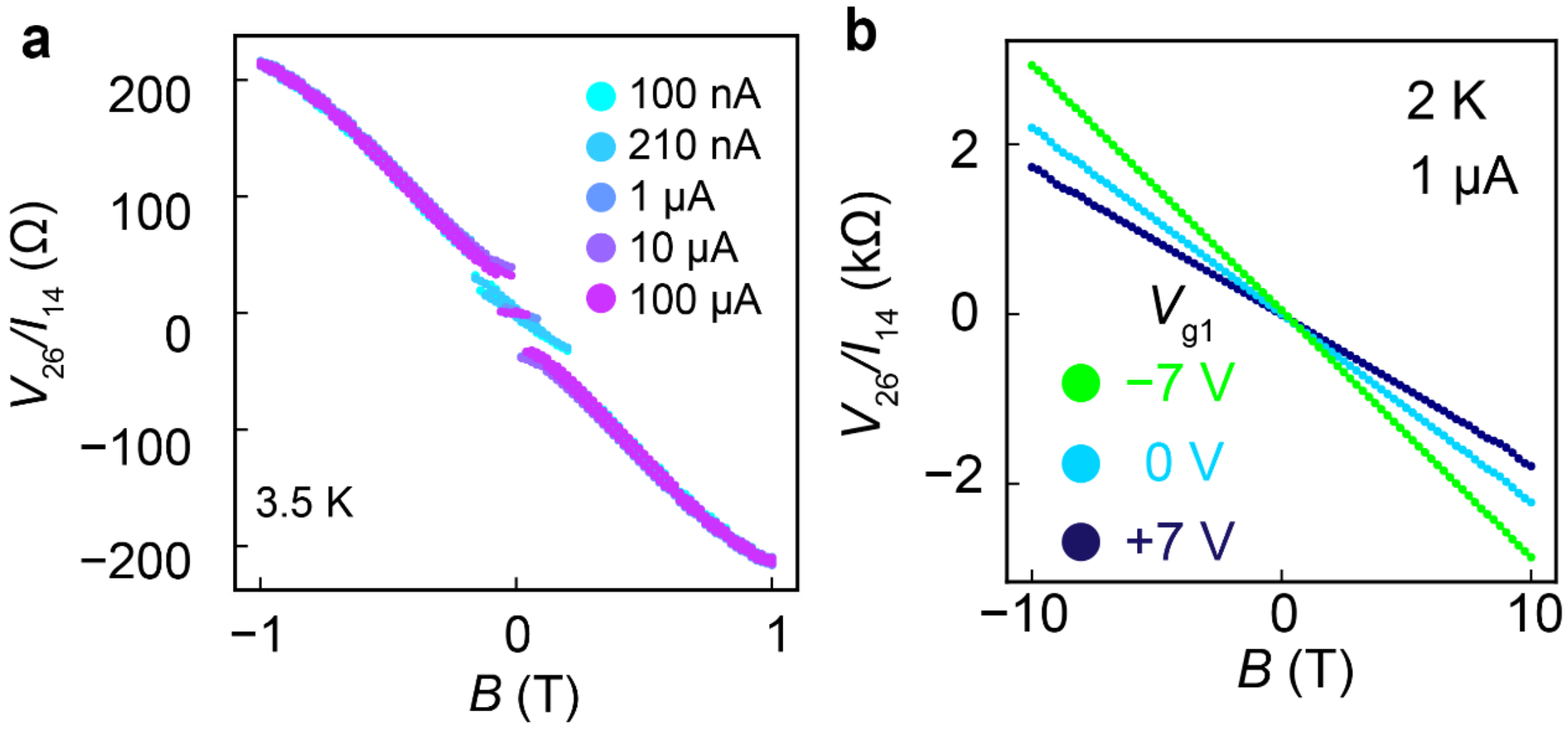

**Supplementary Fig. 9| Current dependence of the Hall resistance a**, Current dependence of the Hall resistance ($V_{26}/I_{14}$) of device D1 measured by the lock-in technique with 5261 Hz at 3.5 K. **b**, Hall resistance ($V_{26}/I_{14}$) *vs*. perpendicular magnetic field $B$ of device D2 measured at various gate voltage $V_{g1}$ (applied to gate G1) at 2 K with 1 μA. As shown in Fig. 3b, the OMR changes its polarity by switching $V_{g1}$ from +7 V to −7 V. However, the Hall resistance does not exhibit the sign change in this $V_{g1}$ region.

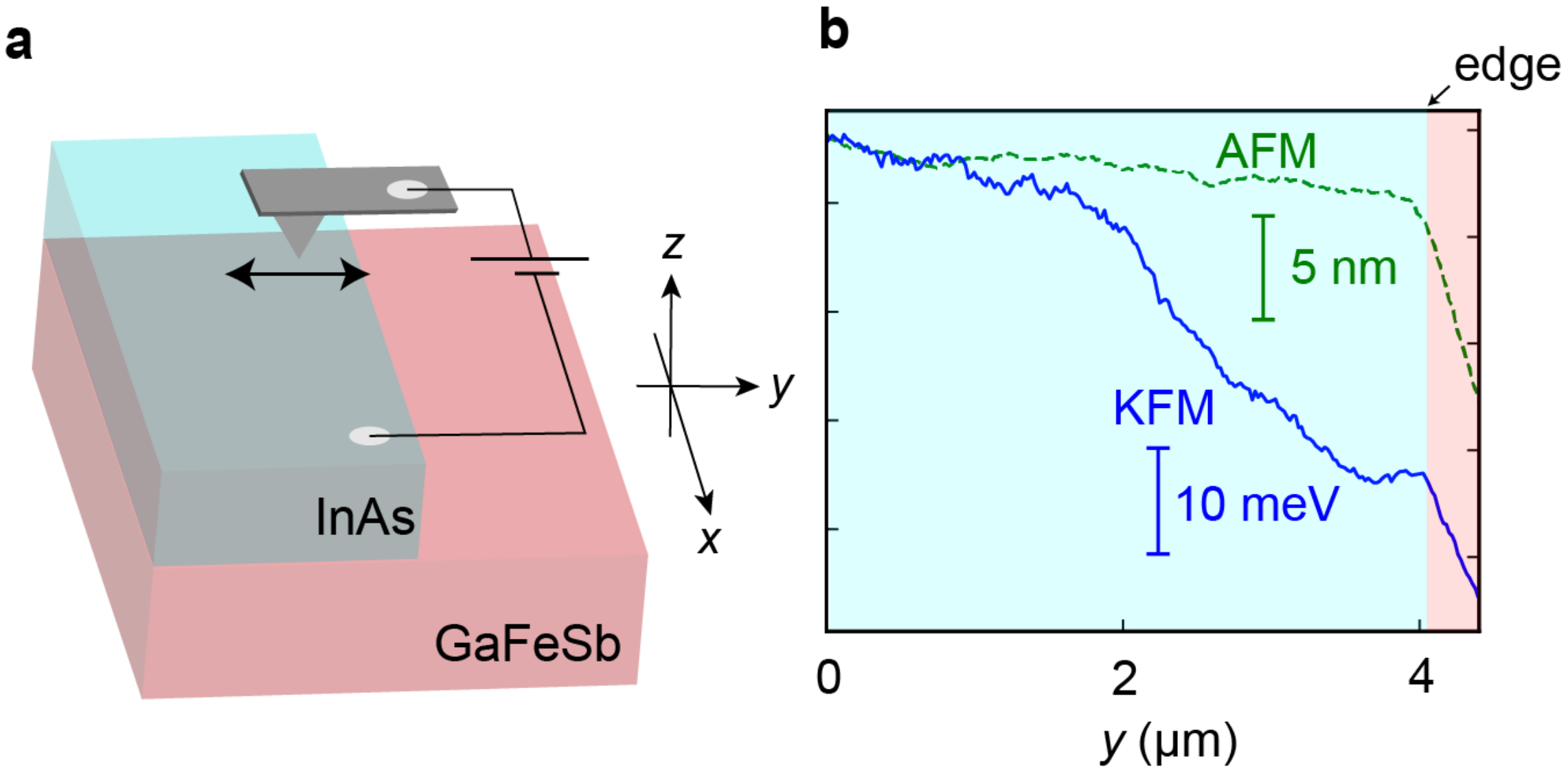

**Supplementary Fig. 10| Kelvin force microscopy (KFM) measurement of InAs/(Ga,Fe)Sb a**, Schematic image of the configuration of the KFM measurement. **b**, KFM (blue solid line) and AFM (green dashed line) results in the *y* direction sweep. The red shaded area is the place where the tip goes through the edge of the sample.

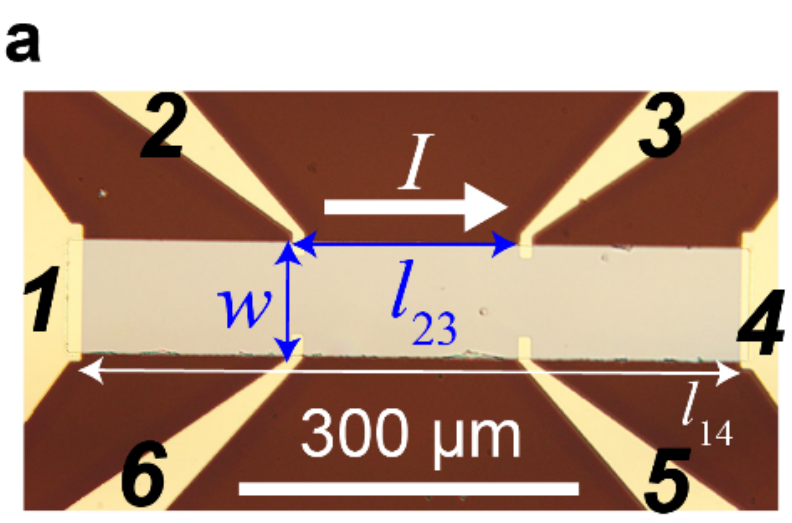

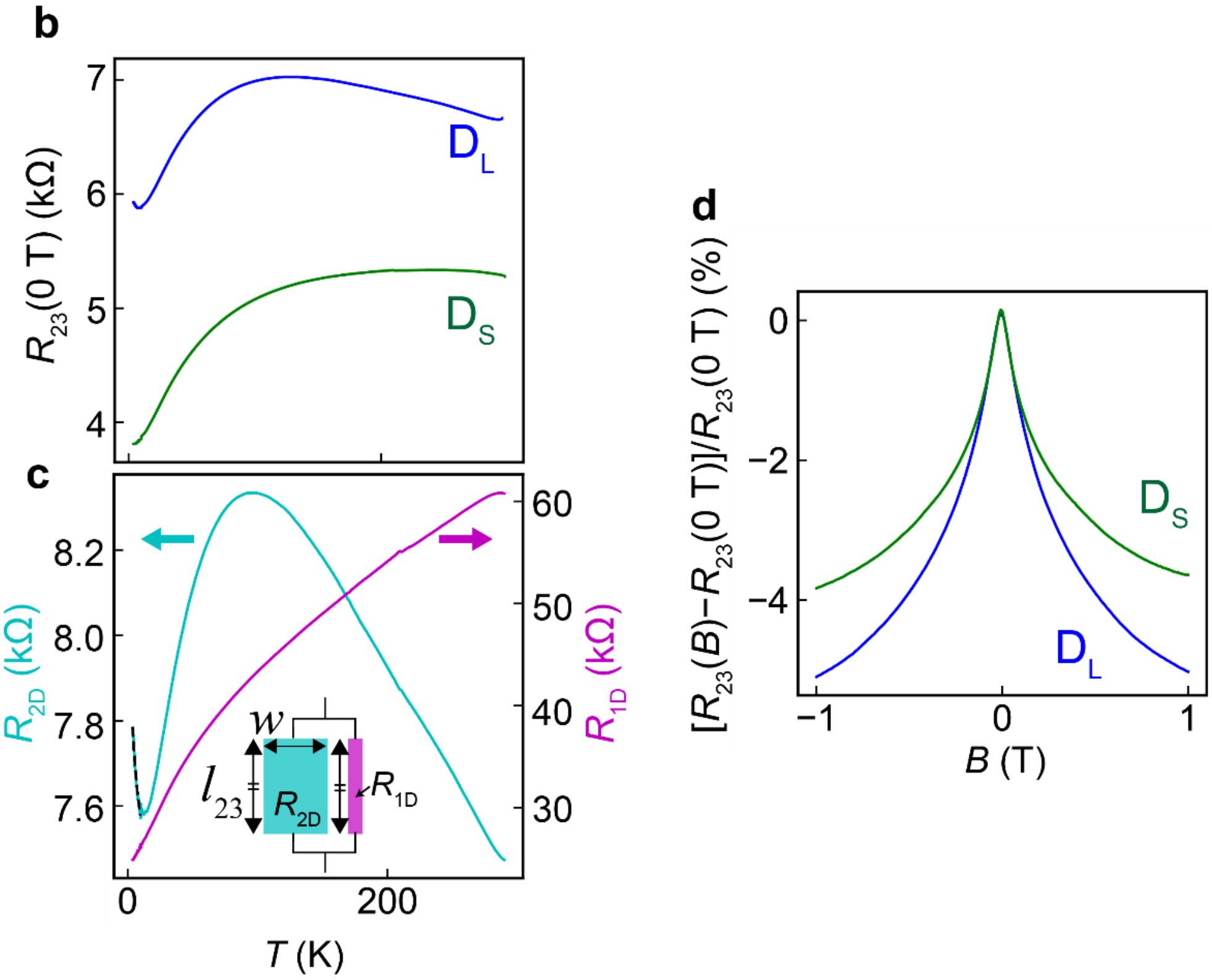

**Supplementary Fig. 11| Device-size effect on the 1D and 2D transport. a**, Optical microscope image (same as Fig. 1b in the main manuscript) of the Hall bar device $D_L$. Here, $w$ and $l_{23}$ indicate the width and the distance between "2" and "3" electrodes, respectively. **b**, Temperature ($T$) dependence of $R_{23}$(0 T) of device $D_L$ (blue) and device $D_S$ (green). **c**, $T$ dependence of $R_{2D}$ (cyan) and $R_{1D}$ (purple) in $D_L$. The inset shows the schematic resistor network representing $R_{23}$. The 2D resistance $R_{2D}$ has the width $w$ and length $l_{23}$, and the 1D resistance $R_{1D}$ has the same length. The black dashed line at $T <$ ~10 K is the fitting result using a logarithmic function, which is characteristic of the Kondo effect ($R_{2D} = R_{c0} - R_{c1} \ln T$, where $R_{c0}$ and $R_{c1}$ are fitting parameters). **d**, Normalized magnetoresistance by the zero-field resistance, $[R_{23}(B)−R_{23}(0\ \mathrm{T})] / R_{23}(0\ \mathrm{T})$, of $D_L$ (blue) and $D_S$ (green).

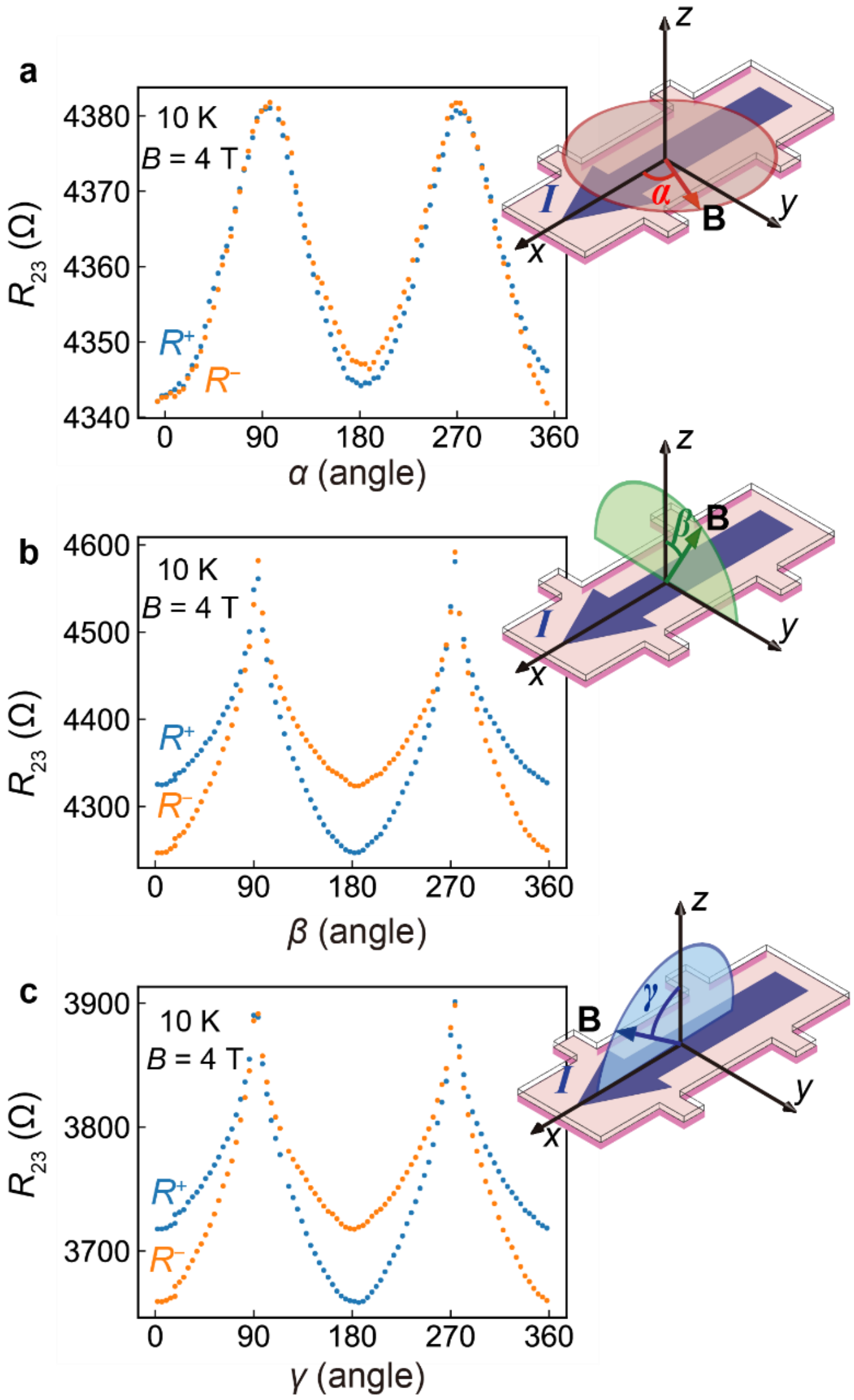

**Supplementary Fig. 12| Angle dependence of the OMR in InAs/(Ga,Fe)Sb a, b,** and **c** Magnetic-field angle dependence of $R_{23}$ (=$V_{23}/I_{14}$) in *xy*, *yz* and *zx* rotation. As shown in the schematic image, each rotation angle is defined as $\alpha$, $\beta$, and $\gamma$ in the *xy*, *yz* and *zx* plane, respectively. Blue and orange dots indicate $R^+$ and $R^-$, respectively, where $R^+$ and $R^-$ are defined as $R_{23}$ when the magnetic field $B$ is positive and negative, respectively. The difference between $R^+$ and $R^-$ corresponds to the OMR magnitude.

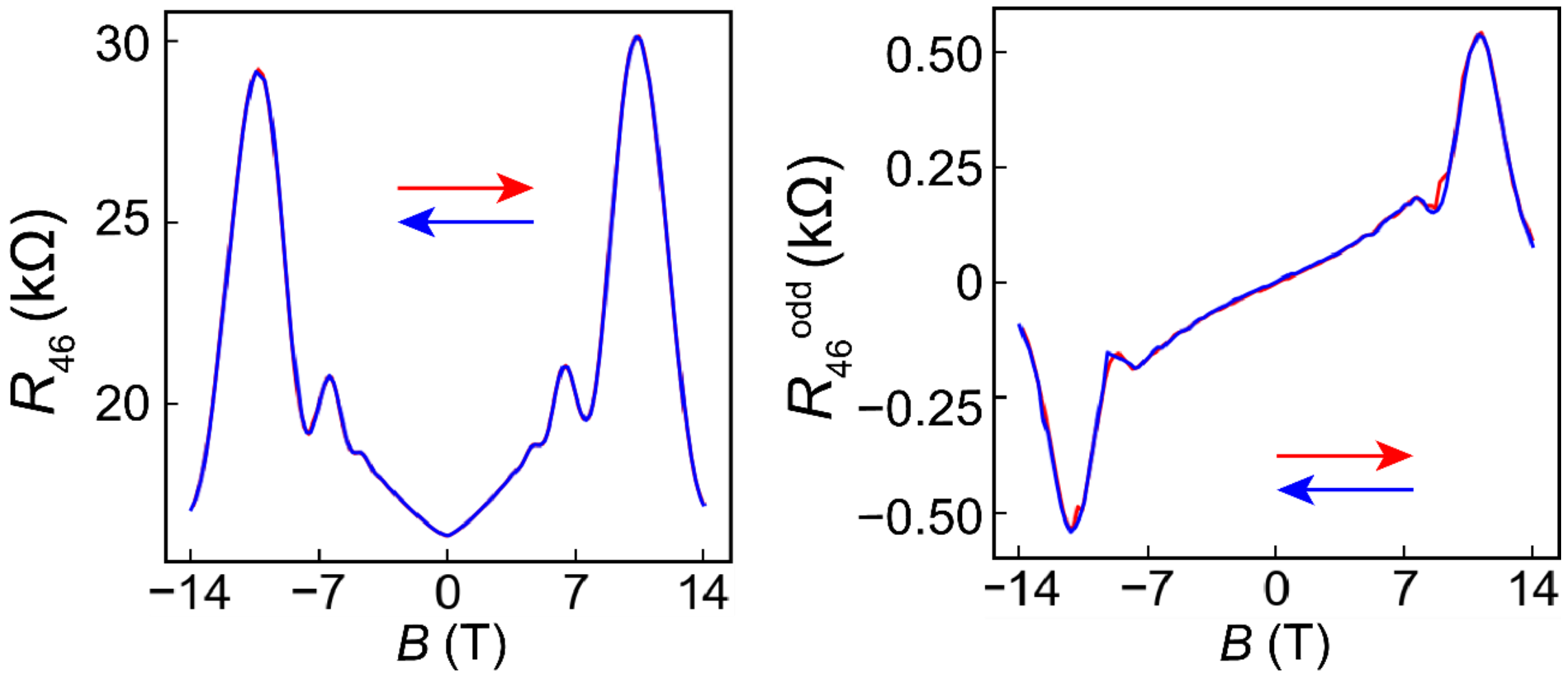

**Supplementary Fig. 13| Magnetoresistance of a nonmagnetic InAs/GaSb bilayer (left side graph) and its odd component (right side graph) at 2 K.** The red and blue arrows indicate sweep direction of the magnetic field. The OMR magnitude is less than 1.8% at 14 T, much smaller than that (13.5% at 10 T) in the InAs/(Ga,Fe)Sb bilayer.

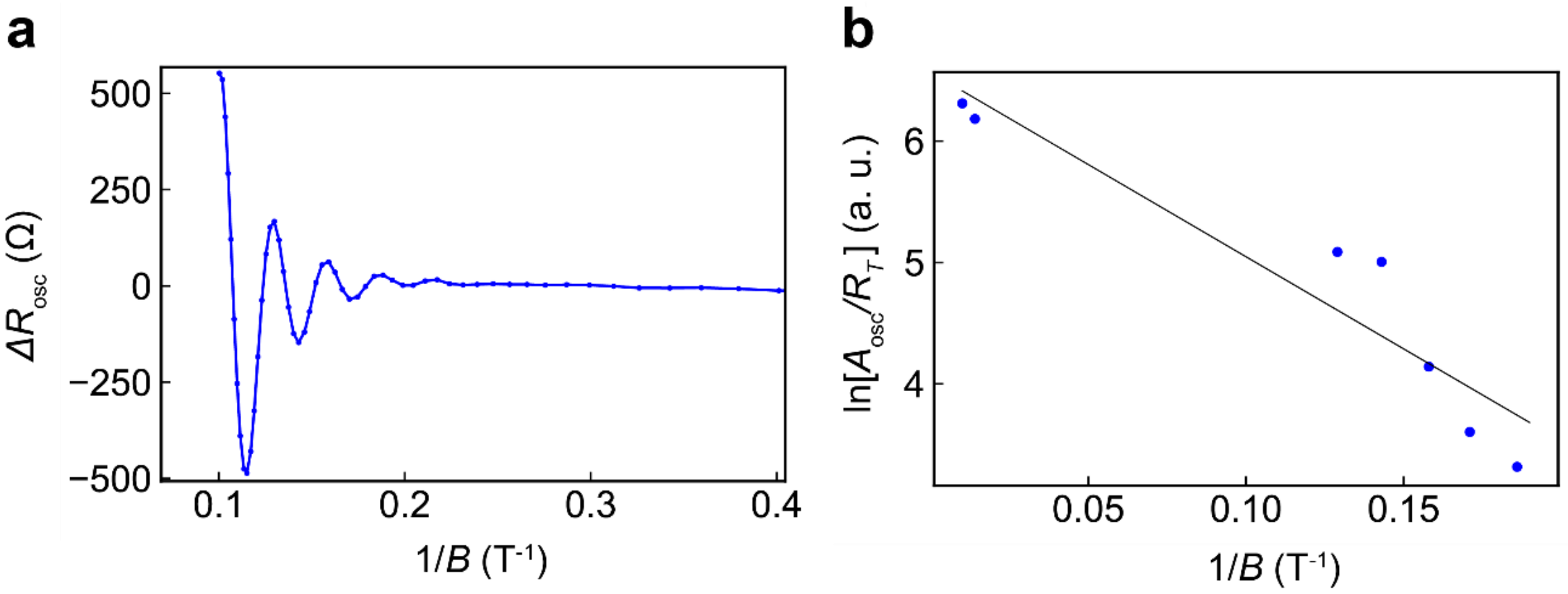

**Supplementary Fig. 14| Dingle plot for sample A. a**, Oscillating component $\Delta R_{osc}$ of sample A extracted from the magnetotransport data shown in Fig. 1c in the main manuscript. The background signal is fitted by a third polynomial function and subtracted from the raw data. **b**, Dingle plot from the data in **a**. $A_{osc}$ is the peak value of the SdH oscillation. $R_T$ is the temperature reduction factor: $R_T = \sinh X / X, X = 2\pi k_B T / \hbar\omega_c$. Here, $k_B$ is Boltzmann's constant, and $\omega_c$ (= $eB/m^*$) is the cyclotron angular frequency. The black line indicates the fitting of the Dingle plot.

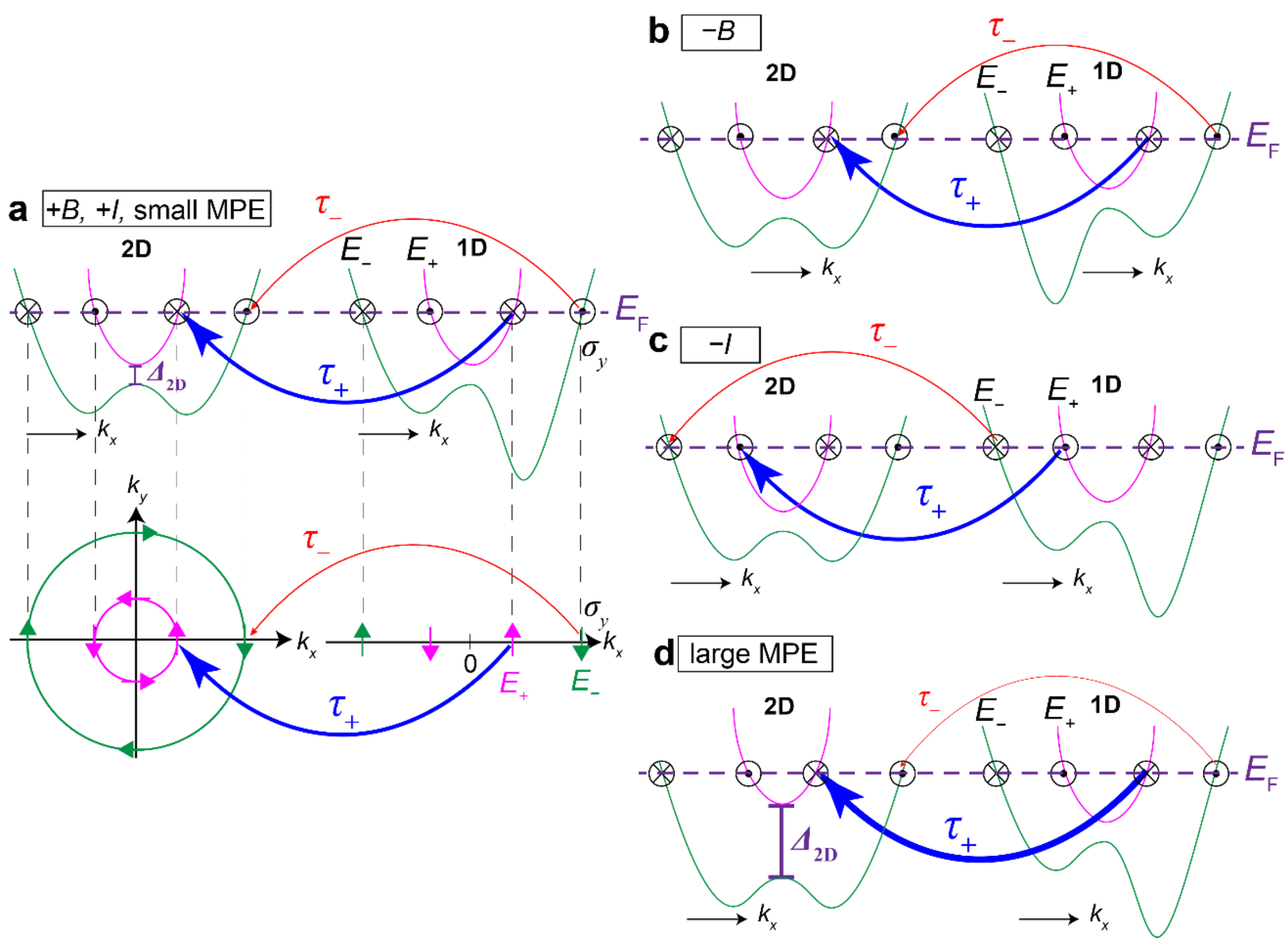

**Supplementary Fig. 15| Mechanism of asymmetric scattering.** **a**, Schematic energy band dispersions (upper panel) and their Fermi surfaces (lower panel) in the $k_x$ direction of the 1D edge (right) and 2D (left) channels in the InAs layer, where a magnetic field $B$ is applied in the $z$ direction perpendicular to the plane. The horizontal purple dashed line indicates the Fermi level $E_F$. Here we consider that the scattering between the 1D and 2D channels is only allowed within the same chirality ($\sigma_y$), which are indicated by the blue and red arrows. These scattering processes have different relaxation times of $\tau_+$ and $\tau_-$, respectively, because of the different density of states between $E_+$ and $E_-$ at $E_F$. **b**, Energy band dispersions when the magnetic field $B$ is reversed. The chirality of $E_+$ and $E_-$, which is determined by the Rashba SOI, is unchanged. Therefore, $\alpha$ (= $\tau_+/\tau_-$) remains unchanged. **c**, When we flip the current $I$, the scattering occurs in the $-k_x$ region. In this case, $\alpha$ (= $\tau_+/\tau_-$) also remains unchanged, thus the OMR in the InAs/(Ga,Fe)Sb bilayer does not depend on the current direction. **d**, The MPE opens a gap (= $\Delta_{2D}$) in the 2D channel. This affects the DOS of each chirality in the 2D channel, which enhances the imbalance of $\tau_+$ and $\tau_-$ and leads to larger OMR.